\documentclass[reprint,amsmath,amssymb,aps,floatfix,superscriptaddress,showpacs,prl]{revtex4-1}

\usepackage{graphicx}
\usepackage{dcolumn}
\usepackage{bm}
\usepackage{color}

\begin{document}

\title{Extended Measurement of the Cosmic-Ray Electron and Positron Spectrum\\ 
 from 11 GeV to 4.8 TeV with the Calorimetric Electron Telescope\\
 on the International Space Station}

\author{O.~Adriani}
\affiliation{Department of Physics, University of Florence, Via Sansone, 1 - 50019 Sesto, Fiorentino, Italy}
\affiliation{INFN Sezione di Florence, Via Sansone, 1 - 50019 Sesto, Fiorentino, Italy}
\author{Y.~Akaike}
\affiliation{Department of Physics, University of Maryland, Baltimore County, 1000 Hilltop Circle, Baltimore, MD 21250, USA}
\affiliation{Astroparticle Physics Laboratory, NASA/GSFC, Greenbelt, MD 20771, USA}
\author{K.~Asano}
\affiliation{Institute for Cosmic Ray Research, The University of Tokyo, 5-1-5 Kashiwa-no-Ha, Kashiwa, Chiba 277-8582, Japan}
\author{Y.~Asaoka}
\email[]{yoichi.asaoka@aoni.waseda.jp}
\affiliation{Research Institute for Science and Engineering, Waseda University, 3-4-1 Okubo, Shinjuku, Tokyo 169-8555, Japan}
\affiliation{JEM Utilization Center, Human Spaceflight Technology Directorate, Japan Aerospace Exploration Agency, 2-1-1 Sengen, Tsukuba, Ibaraki 305-8505, Japan}
\author{M.G.~Bagliesi}
\affiliation{Department of Physical Sciences, Earth and Environment, University of Siena, via Roma 56, 53100 Siena, Italy}
\affiliation{INFN Sezione di Pisa, Polo Fibonacci, Largo B. Pontecorvo, 3 - 56127 Pisa, Italy}
\author{E.~Berti} 
\affiliation{Department of Physics, University of Florence, Via Sansone, 1 - 50019 Sesto, Fiorentino, Italy}
\affiliation{INFN Sezione di Florence, Via Sansone, 1 - 50019 Sesto, Fiorentino, Italy}
\author{G.~Bigongiari}
\affiliation{Department of Physical Sciences, Earth and Environment, University of Siena, via Roma 56, 53100 Siena, Italy}
\affiliation{INFN Sezione di Pisa, Polo Fibonacci, Largo B. Pontecorvo, 3 - 56127 Pisa, Italy}
\author{W.R.~Binns}
\affiliation{Department of Physics, Washington University, One Brookings Drive, St. Louis, MO 63130-4899, USA}
\author{S.~Bonechi}
\affiliation{Department of Physical Sciences, Earth and Environment, University of Siena, via Roma 56, 53100 Siena, Italy}
\affiliation{INFN Sezione di Pisa, Polo Fibonacci, Largo B. Pontecorvo, 3 - 56127 Pisa, Italy}
\author{M.~Bongi}
\affiliation{Department of Physics, University of Florence, Via Sansone, 1 - 50019 Sesto, Fiorentino, Italy}
\affiliation{INFN Sezione di Florence, Via Sansone, 1 - 50019 Sesto, Fiorentino, Italy}
\author{P.~Brogi}
\affiliation{Department of Physical Sciences, Earth and Environment, University of Siena, via Roma 56, 53100 Siena, Italy}
\affiliation{INFN Sezione di Pisa, Polo Fibonacci, Largo B. Pontecorvo, 3 - 56127 Pisa, Italy}
\author{J.H.~Buckley}
\affiliation{Department of Physics, Washington University, One Brookings Drive, St. Louis, MO 63130-4899, USA}
\author{N.~Cannady}
\affiliation{Department of Physics and Astronomy, Louisiana State University, 202 Nicholson Hall, Baton Rouge, LA 70803, USA}
\author{G.~Castellini}
\affiliation{Institute of Applied Physics (IFAC),  National Research Council (CNR), Via Madonna del Piano, 10, 50019 Sesto, Fiorentino, Italy}
\author{C.~Checchia}
\affiliation{Department of Physics and Astronomy, University of Padova, Via Marzolo, 8, 35131 Padova, Italy}
\affiliation{INFN Sezione di Padova, Via Marzolo, 8, 35131 Padova, Italy} 
\author{M.L.~Cherry}
\affiliation{Department of Physics and Astronomy, Louisiana State University, 202 Nicholson Hall, Baton Rouge, LA 70803, USA}
\author{G.~Collazuol}
\affiliation{Department of Physics and Astronomy, University of Padova, Via Marzolo, 8, 35131 Padova, Italy}
\affiliation{INFN Sezione di Padova, Via Marzolo, 8, 35131 Padova, Italy} 
\author{V.~Di~Felice}
\affiliation{University of Rome ``Tor Vergata'', Via della Ricerca Scientifica 1, 00133 Rome, Italy}
\affiliation{INFN Sezione di Rome ``Tor Vergata'', Via della Ricerca Scientifica 1, 00133 Rome, Italy}
\author{K.~Ebisawa}
\affiliation{Institute of Space and Astronautical Science, Japan Aerospace Exploration Agency, 3-1-1 Yoshinodai, Chuo, Sagamihara, Kanagawa 252-5210, Japan}
\author{H.~Fuke}
\affiliation{Institute of Space and Astronautical Science, Japan Aerospace Exploration Agency, 3-1-1 Yoshinodai, Chuo, Sagamihara, Kanagawa 252-5210, Japan}
\author{T.G.~Guzik}
\affiliation{Department of Physics and Astronomy, Louisiana State University, 202 Nicholson Hall, Baton Rouge, LA 70803, USA}
\author{T.~Hams}
\affiliation{Department of Physics, University of Maryland, Baltimore County, 1000 Hilltop Circle, Baltimore, MD 21250, USA}
\affiliation{CRESST and Astroparticle Physics Laboratory NASA/GSFC, Greenbelt, MD 20771, USA}
\author{M.~Hareyama}
\affiliation{St. Marianna University School of Medicine, 2-16-1, Sugao, Miyamae-ku, Kawasaki, Kanagawa 216-8511, Japan}
\author{N.~Hasebe}
\affiliation{Research Institute for Science and Engineering, Waseda University, 3-4-1 Okubo, Shinjuku, Tokyo 169-8555, Japan}
\author{K.~Hibino}
\affiliation{Kanagawa University, 3-27-1 Rokkakubashi, Kanagawa, Yokohama, Kanagawa 221-8686, Japan}
\author{M.~Ichimura}
\affiliation{Faculty of Science and Technology, Graduate School of Science and Technology, Hirosaki University, 3, Bunkyo, Hirosaki, Aomori 036-8561, Japan}
\author{K.~Ioka}
\affiliation{Yukawa Institute for Theoretical Physics, Kyoto University, Kitashirakawa Oiwakecho, Sakyo, Kyoto 606-8502, Japan}
\author{W.~Ishizaki}
\affiliation{Institute for Cosmic Ray Research, The University of Tokyo, 5-1-5 Kashiwa-no-Ha, Kashiwa, Chiba 277-8582, Japan}
\author{M.H.~Israel}
\affiliation{Department of Physics, Washington University, One Brookings Drive, St. Louis, MO 63130-4899, USA}
\author{K.~Kasahara}
\affiliation{Research Institute for Science and Engineering, Waseda University, 3-4-1 Okubo, Shinjuku, Tokyo 169-8555, Japan}
\author{J.~Kataoka}
\affiliation{Research Institute for Science and Engineering, Waseda University, 3-4-1 Okubo, Shinjuku, Tokyo 169-8555, Japan}
\author{R.~Kataoka}
\affiliation{National Institute of Polar Research, 10-3, Midori-cho, Tachikawa, Tokyo 190-8518, Japan}
\author{Y.~Katayose}
\affiliation{Faculty of Engineering, Division of Intelligent Systems Engineering, Yokohama National University, 79-5 Tokiwadai, Hodogaya, Yokohama 240-8501, Japan}
\author{C.~Kato}
\affiliation{Faculty of Science, Shinshu University, 3-1-1 Asahi, Matsumoto, Nagano 390-8621, Japan}
\author{N.~Kawanaka}
\affiliation{Hakubi Center, Kyoto University, Yoshida Honmachi, Sakyo-ku, Kyoto, 606-8501, Japan}
\affiliation{Department of Astronomy, Graduate School of Science, Kyoto University, Kitashirakawa Oiwake-cho, Sakyo-ku, Kyoto, 606-8502, Japan}
\author{Y.~Kawakubo}
\affiliation{College of Science and Engineering, Department of Physics and Mathematics, Aoyama Gakuin University,  5-10-1 Fuchinobe, Chuo, Sagamihara, Kanagawa 252-5258, Japan}
\author{K.~Kohri} 
\affiliation{Institute of Particle and Nuclear Studies, High Energy Accelerator Research Organization, 1-1 Oho, Tsukuba, Ibaraki, 305-0801, Japan} 
\author{H.S.~Krawczynski}
\affiliation{Department of Physics, Washington University, One Brookings Drive, St. Louis, MO 63130-4899, USA}
\author{J.F.~Krizmanic}
\affiliation{CRESST and Astroparticle Physics Laboratory NASA/GSFC, Greenbelt, MD 20771, USA}
\affiliation{Department of Physics, University of Maryland, Baltimore County, 1000 Hilltop Circle, Baltimore, MD 21250, USA}
\author{T.~Lomtadze}
\affiliation{INFN Sezione di Pisa, Polo Fibonacci, Largo B. Pontecorvo, 3 - 56127 Pisa, Italy}
\author{P.~Maestro}
\affiliation{Department of Physical Sciences, Earth and Environment, University of Siena, via Roma 56, 53100 Siena, Italy}
\affiliation{INFN Sezione di Pisa, Polo Fibonacci, Largo B. Pontecorvo, 3 - 56127 Pisa, Italy}
\author{P.S.~Marrocchesi}
\affiliation{Department of Physical Sciences, Earth and Environment, University of Siena, via Roma 56, 53100 Siena, Italy}
\affiliation{INFN Sezione di Pisa, Polo Fibonacci, Largo B. Pontecorvo, 3 - 56127 Pisa, Italy}
\author{A.M.~Messineo}
\affiliation{University of Pisa, Polo Fibonacci, Largo B. Pontecorvo, 3 - 56127 Pisa, Italy}
\affiliation{INFN Sezione di Pisa, Polo Fibonacci, Largo B. Pontecorvo, 3 - 56127 Pisa, Italy}
\author{J.W.~Mitchell}
\affiliation{Astroparticle Physics Laboratory, NASA/GSFC, Greenbelt, MD 20771, USA}
\author{S.~Miyake}
\affiliation{Department of Electrical and Electronic Systems Engineering, National Institute of Technology, Ibaraki College, 866 Nakane, Hitachinaka, Ibaraki 312-8508 Japan}
\author{A.A.~Moiseev}
\affiliation{Department of Astronomy, University of Maryland, College Park, Maryland 20742, USA }
\affiliation{CRESST and Astroparticle Physics Laboratory NASA/GSFC, Greenbelt, MD 20771, USA}
\author{K.~Mori}
\affiliation{Research Institute for Science and Engineering, Waseda University, 3-4-1 Okubo, Shinjuku, Tokyo 169-8555, Japan}
\affiliation{Institute of Space and Astronautical Science, Japan Aerospace Exploration Agency, 3-1-1 Yoshinodai, Chuo, Sagamihara, Kanagawa 252-5210, Japan}
\author{M.~Mori}
\affiliation{Department of Physical Sciences, College of Science and Engineering, Ritsumeikan University, Shiga 525-8577, Japan}
\author{N.~Mori}
\affiliation{INFN Sezione di Florence, Via Sansone, 1 - 50019 Sesto, Fiorentino, Italy}
\author{H.M.~Motz}
\affiliation{International Center for Science and Engineering Programs, Waseda University, 3-4-1 Okubo, Shinjuku, Tokyo 169-8555, Japan}
\author{K.~Munakata}
\affiliation{Faculty of Science, Shinshu University, 3-1-1 Asahi, Matsumoto, Nagano 390-8621, Japan}
\author{H.~Murakami}
\affiliation{Research Institute for Science and Engineering, Waseda University, 3-4-1 Okubo, Shinjuku, Tokyo 169-8555, Japan}
\author{S.~Nakahira}
\affiliation{RIKEN, 2-1 Hirosawa, Wako, Saitama 351-0198, Japan}
\author{J.~Nishimura}
\affiliation{Institute of Space and Astronautical Science, Japan Aerospace Exploration Agency, 3-1-1 Yoshinodai, Chuo, Sagamihara, Kanagawa 252-5210, Japan}
\author{G.A.~de~Nolfo}
\affiliation{Heliospheric Physics Laboratory, NASA/GSFC, Greenbelt, MD 20771, USA}
\author{S.~Okuno}
\affiliation{Kanagawa University, 3-27-1 Rokkakubashi, Kanagawa, Yokohama, Kanagawa 221-8686, Japan}
\author{J.F.~Ormes}
\affiliation{Department of Physics and Astronomy, University of Denver, Physics Building, Room 211, 2112 East Wesley Ave., Denver, CO 80208-6900, USA}
\author{S.~Ozawa}
\affiliation{Research Institute for Science and Engineering, Waseda University, 3-4-1 Okubo, Shinjuku, Tokyo 169-8555, Japan}
\author{L.~Pacini}
\affiliation{Department of Physics, University of Florence, Via Sansone, 1 - 50019 Sesto, Fiorentino, Italy}
\affiliation{Institute of Applied Physics (IFAC),  National Research Council (CNR), Via Madonna del Piano, 10, 50019 Sesto, Fiorentino, Italy}
\affiliation{INFN Sezione di Florence, Via Sansone, 1 - 50019 Sesto, Fiorentino, Italy}
\author{F.~Palma}
\affiliation{University of Rome ``Tor Vergata'', Via della Ricerca Scientifica 1, 00133 Rome, Italy}
\affiliation{INFN Sezione di Rome ``Tor Vergata'', Via della Ricerca Scientifica 1, 00133 Rome, Italy}
\author{P.~Papini}
\affiliation{INFN Sezione di Florence, Via Sansone, 1 - 50019 Sesto, Fiorentino, Italy}
\author{A.V.~Penacchioni}
\affiliation{Department of Physical Sciences, Earth and Environment, University of Siena, via Roma 56, 53100 Siena, Italy}
\affiliation{ASI Science Data Center (ASDC), Via del Politecnico snc, 00133 Rome, Italy}
\author{B.F.~Rauch}
\affiliation{Department of Physics, Washington University, One Brookings Drive, St. Louis, MO 63130-4899, USA}
\author{S.B.~Ricciarini}
\affiliation{Institute of Applied Physics (IFAC),  National Research Council (CNR), Via Madonna del Piano, 10, 50019 Sesto, Fiorentino, Italy}
\affiliation{INFN Sezione di Florence, Via Sansone, 1 - 50019 Sesto, Fiorentino, Italy}
\author{K.~Sakai}
\affiliation{CRESST and Astroparticle Physics Laboratory NASA/GSFC, Greenbelt, MD 20771, USA}
\affiliation{Department of Physics, University of Maryland, Baltimore County, 1000 Hilltop Circle, Baltimore, MD 21250, USA}
\author{T.~Sakamoto}
\affiliation{College of Science and Engineering, Department of Physics and Mathematics, Aoyama Gakuin University,  5-10-1 Fuchinobe, Chuo, Sagamihara, Kanagawa 252-5258, Japan}
\author{M.~Sasaki}
\affiliation{CRESST and Astroparticle Physics Laboratory NASA/GSFC, Greenbelt, MD 20771, USA}
\affiliation{Department of Astronomy, University of Maryland, College Park, Maryland 20742, USA }
\author{Y.~Shimizu}
\affiliation{Kanagawa University, 3-27-1 Rokkakubashi, Kanagawa, Yokohama, Kanagawa 221-8686, Japan}
\author{A.~Shiomi}
\affiliation{College of Industrial Technology, Nihon University, 1-2-1 Izumi, Narashino, Chiba 275-8575, Japan}
\author{R.~Sparvoli}
\affiliation{University of Rome ``Tor Vergata'', Via della Ricerca Scientifica 1, 00133 Rome, Italy}
\affiliation{INFN Sezione di Rome ``Tor Vergata'', Via della Ricerca Scientifica 1, 00133 Rome, Italy}
\author{P.~Spillantini}
\affiliation{Department of Physics, University of Florence, Via Sansone, 1 - 50019 Sesto, Fiorentino, Italy}
\author{F.~Stolzi}
\affiliation{Department of Physical Sciences, Earth and Environment, University of Siena, via Roma 56, 53100 Siena, Italy}
\affiliation{INFN Sezione di Pisa, Polo Fibonacci, Largo B. Pontecorvo, 3 - 56127 Pisa, Italy}
\author{J.E.~Suh} 
\affiliation{Department of Physical Sciences, Earth and Environment, University of Siena, via Roma 56, 53100 Siena, Italy}
\affiliation{INFN Sezione di Pisa, Polo Fibonacci, Largo B. Pontecorvo, 3 - 56127 Pisa, Italy}
\author{A.~Sulaj} 
\affiliation{Department of Physical Sciences, Earth and Environment, University of Siena, via Roma 56, 53100 Siena, Italy}
\affiliation{INFN Sezione di Pisa, Polo Fibonacci, Largo B. Pontecorvo, 3 - 56127 Pisa, Italy}
\author{I.~Takahashi}
\affiliation{Kavli Institute for the Physics and Mathematics of the Universe, The University of Tokyo, 5-1-5 Kashiwanoha, Kashiwa, 277-8583, Japan}
\author{M.~Takayanagi}
\affiliation{Institute of Space and Astronautical Science, Japan Aerospace Exploration Agency, 3-1-1 Yoshinodai, Chuo, Sagamihara, Kanagawa 252-5210, Japan}
\author{M.~Takita}
\affiliation{Institute for Cosmic Ray Research, The University of Tokyo, 5-1-5 Kashiwa-no-Ha, Kashiwa, Chiba 277-8582, Japan}
\author{T.~Tamura}
\affiliation{Kanagawa University, 3-27-1 Rokkakubashi, Kanagawa, Yokohama, Kanagawa 221-8686, Japan}
\author{N.~Tateyama}
\affiliation{Kanagawa University, 3-27-1 Rokkakubashi, Kanagawa, Yokohama, Kanagawa 221-8686, Japan}
\author{T.~Terasawa}
\affiliation{RIKEN, 2-1 Hirosawa, Wako, Saitama 351-0198, Japan}
\author{H.~Tomida}
\affiliation{Institute of Space and Astronautical Science, Japan Aerospace Exploration Agency, 3-1-1 Yoshinodai, Chuo, Sagamihara, Kanagawa 252-5210, Japan}
\author{S.~Torii}
\email[]{torii.shoji@waseda.jp}
\affiliation{Research Institute for Science and Engineering, Waseda University, 3-4-1 Okubo, Shinjuku, Tokyo 169-8555, Japan}
\affiliation{JEM Utilization Center, Human Spaceflight Technology Directorate, Japan Aerospace Exploration Agency, 2-1-1 Sengen, Tsukuba, Ibaraki 305-8505, Japan}
\affiliation{School of Advanced Science and Engineering, Waseda University, 3-4-1 Okubo, Shinjuku, Tokyo 169-8555, Japan}
\author{Y.~Tsunesada}
\affiliation{Division of Mathematics and Physics, Graduate School of Science, Osaka City University, 3-3-138 Sugimoto, Sumiyoshi, Osaka 558-8585, Japan}
\author{Y.~Uchihori}
\affiliation{National Institutes for Quantum and Radiation Science and Technology, 4-9-1 Anagawa, Inage, Chiba 263-8555, JAPAN}
\author{S.~Ueno}
\affiliation{Institute of Space and Astronautical Science, Japan Aerospace Exploration Agency, 3-1-1 Yoshinodai, Chuo, Sagamihara, Kanagawa 252-5210, Japan}
\author{E.~Vannuccini}
\affiliation{INFN Sezione di Florence, Via Sansone, 1 - 50019 Sesto, Fiorentino, Italy}
\author{J.P.~Wefel}
\affiliation{Department of Physics and Astronomy, Louisiana State University, 202 Nicholson Hall, Baton Rouge, LA 70803, USA}
\author{K.~Yamaoka}
\affiliation{Nagoya University, Furo, Chikusa, Nagoya 464-8601, Japan}
\author{S.~Yanagita}
\affiliation{College of Science, Ibaraki University, 2-1-1 Bunkyo, Mito, Ibaraki 310-8512, Japan}
\author{A.~Yoshida}
\affiliation{College of Science and Engineering, Department of Physics and Mathematics, Aoyama Gakuin University,  5-10-1 Fuchinobe, Chuo, Sagamihara, Kanagawa 252-5258, Japan}
\author{K.~Yoshida}
\affiliation{Department of Electronic Information Systems, Shibaura Institute of Technology, 307 Fukasaku, Minuma, Saitama 337-8570, Japan}

\collaboration{CALET Collaboration}

\date{\today}

\begin{abstract}
Extended results on the cosmic-ray electron + positron spectrum from 11 GeV to 4.8 TeV are presented based on observations with the Calorimetric Electron Telescope (CALET) on the International Space Station utilizing the data up to November 2017. 
The analysis uses the full detector acceptance at high energies, approximately doubling the statistics compared to the previous result. 
CALET is an all-calorimetric instrument with a total thickness of 30 $X_0$ at normal incidence and fine imaging capability, designed to achieve large proton rejection and excellent energy resolution  well into the TeV energy region.  
The observed energy spectrum in the region below 1~TeV  shows good agreement with Alpha Magnetic Spectrometer (AMS-02) data. 
In the energy region below $\sim$300~GeV, CALET's spectral index is found to be consistent with the AMS-02, Fermi Large Area Telescope (Fermi-LAT), and DArk Matter Particle Explorer (DAMPE), while from 300 to 600~GeV the spectrum is significantly softer than 
the spectra from the latter two experiments.
The absolute flux of CALET is consistent with other experiments at 
around a few 
tens of GeV. 
However, it is lower than those of DAMPE and Fermi-LAT 
with the difference increasing up to several hundred GeV.
The observed energy spectrum above $\sim$1~TeV suggests a flux suppression consistent 
within the errors
with the 
results of DAMPE, 
while CALET does not observe any significant evidence for a narrow spectral feature in the energy region around 1.4~TeV. 
Our measured all-electron flux, including statistical errors and a detailed breakdown of the systematic errors, is tabulated
in the Supplemental Material 
in order to allow more refined spectral analyses based on our data. 
\end{abstract}


\maketitle

\section{Introduction} 

\begin{figure*}[tbh!]
\begin{center}
\includegraphics[width=\linewidth]{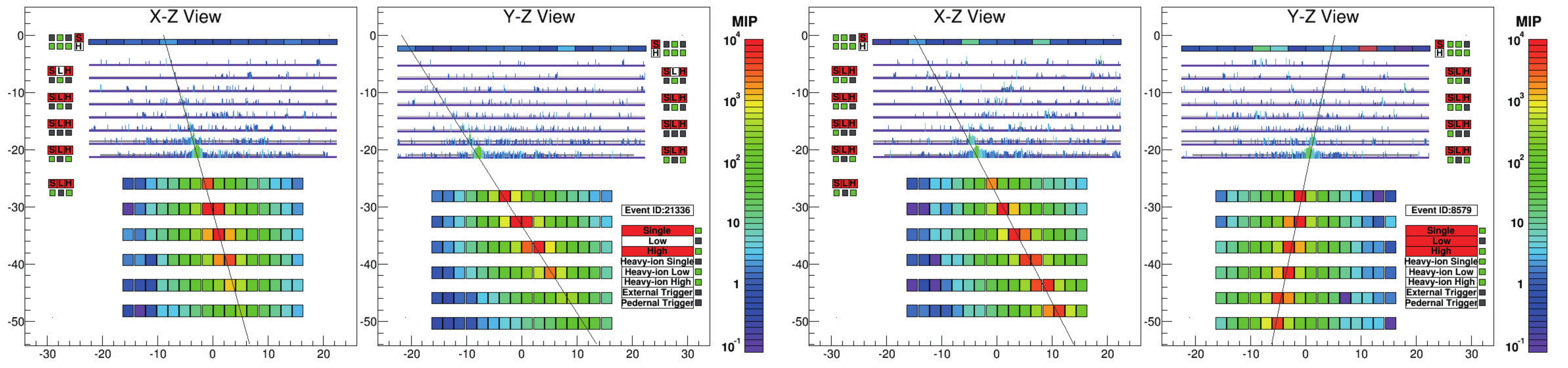}
\caption{Examples of TeV event candidates showing energy deposit in each detector channel in the $X-Z$ and $Y-Z$ views.
(Left) An electron (or positron) candidate (reconstructed energy of 3.05~TeV
and energy deposit sum of 2.89 TeV). 
(Right) A proton candidate (energy deposit sum of 2.89~TeV).
}
\label{fig:tevevt}
\end{center}
\end{figure*}

High-energy cosmic-ray electrons provide a unique probe of nearby cosmic accelerators. 
Electrons rapidly lose energy via inverse Compton scattering and synchrotron emission
during propagation in the Galaxy. 
Since their diffusion distance above 1~TeV is less than 1~kpc, 
only a few potential TeV sources are expected 
in the vicinity of the Solar System. 
A precise measurement of the electron spectrum in the TeV region might reveal 
interesting spectral features to provide the first experimental evidence of 
the possible presence of a nearby cosmic-ray source~\cite{nishimura1980, kobayashi2004}. 
In addition, the prominent increase of the positron fraction over 10~GeV established by 
the Payload for Antimatter Matter Exploration and Light-nuclei Astrophysics (PAMELA) \cite{PAMELA-pe} and the Alpha Magnetic Spectrometer (AMS-02) \cite{AMS02-pe}
may require a primary source component for positrons in addition to the generally accepted secondary origin. 
Candidates for such primary sources range from astrophysical (pulsar) to exotic (dark matter). 
Since these primary sources emit electron-positron pairs, 
it is expected that the all-electron (electrons $+$ positrons) spectrum would exhibit a spectral feature, near the 
highest energy range of the primary component.

The CALET Collaboration managing the CALorimetric Electron Telescope (CALET)~\cite{torii2015},  
a space-based instrument optimized for the measurement 
of the all-electron spectrum, published its first result 
in the energy range from 10 GeV to 3 TeV~\cite{CALET2017}.
Subsequently, the DArk Matter Particle Explorer (DAMPE) collaboration published
their all-electron spectrum in the energy range from 25~GeV to 4.6~TeV~\cite{DAMPE2017}.

In this Letter, we present an updated version of the CALET all-electron spectrum. 
Using 780 days of flight data 
from October 13, 2015 to November 30, 2017 and the full geometrical acceptance in the high-energy region, 
we have increased our statistics by a factor of $\sim$2 compared to Ref.~\cite{CALET2017}. 
The energy range is also extended up to 4.75~TeV. 
Features 
of the spectrum measured by CALET are discussed, 
particularly in relation to the break 
reported 
by DAMPE at 0.9 TeV.
The possible presence of a peak close to 1.4~TeV is tested with CALET data by using exactly the same energy binning as that of DAMPE. 
The systematic uncertainties are classified into several categories in order to allow for more sensitive
interpretative studies using the CALET spectrum. 

\section{CALET Instrument}

CALET employs a fully active calorimeter with 30 radiation-length thickness for particles at normal incidence.
It consists of a charge detector (CHD), 
a 3 radiation-length thick imaging calorimeter (IMC) and a 27 radiation-length thick total 
absorption calorimeter (TASC), 
having a field of view of  $\sim$45$^\circ$ from zenith
and a geometrical factor of $\sim$1040~cm$^2$sr for high-energy electrons.

CHD, which identifies the charge of the incident particle, 
is comprised of 
a
pair of plastic scintillator hodoscopes arranged in two orthogonal layers. 
IMC is a sampling calorimeter 
alternating thin layers of Tungsten absorber, optimized in thickness and position, with layers of scintillating fibers
read-out individually. 
TASC is a tightly packed lead-tungstate (PbWO$_4$; PWO) hodoscope, capable of 
almost complete absorption of the TeV-electron 
showers.
A more complete description of the instrument is given in the Supplemental Material of Ref.~\cite{CALET2017}.

Figure~\ref{fig:tevevt} shows a 3.05~TeV-electron candidate and a proton candidate with
comparable energy deposit 
(2.89~TeV) 
in the detector.
Compared to hadron showers which have significant leakage, 
the containment of the electromagnetic shower creates a 
difference in shower shape especially in the bottom part of TASC, allowing for 
an accurate electron identification in the presence of a large hadron background. 
Together with the precision energy measurements from total absorption of electromagnetic showers,
it is possible to derive the electron spectrum well into the TeV region with a straightforward
and reliable analysis.

The instrument was launched on August 19, 2015 and emplaced on the 
Japanese Experiment Module-Exposed Facility on the International Space Station 
with an expected mission duration of five years (or more). 
Scientific observations~\cite{asaoka2018} were started on October 13, 2015, 
and smooth and continuous operations have taken place since then.

\section{Data Analysis} 
We have analyzed 780 days of flight data collected with a high-energy shower trigger~\cite{asaoka2018}.
Total live time in this period was 15,811 hours, corresponding to a live time fraction of 84\%.
The analysis was extended to use the full detector acceptance at
higher energies as explained further down, otherwise it was done following the
standard analysis procedure described in Ref.~\cite{CALET2017}.

A Monte Carlo (MC) program was used to simulate physics processes and detector response based on the 
simulation package EPICS~\cite{EPICS,EPICSprl} (EPICS9.20 / Cosmos8.00).
Using MC event samples of electrons and protons, 
event selection and event reconstruction efficiencies, energy correction factor, and background contamination
were derived. 
An independent analysis based on Geant4~\cite{Geant4} was performed,
and small differences between the MC models are included in the systematic uncertainties. 
The detector model used in the Geant4 simulation is almost identical to the CALET computer aided design model. 
The Geant4 simulation employs the hadronic interaction models FTFP\_BERT as the physics list, 
while DPMJET3~\cite{dpmjet3prl} is chosen as the hadronic interaction model 
in the EPICS simulation. 

While excellent energy resolution inside the TeV region is 
one of the most important features of a thick calorimeter instrument
like CALET or DAMPE, calibration errors must be carefully assessed 
and taken into account in the estimation of the actual energy resolution.
Our energy calibration~\cite{asaoka2017} includes the evaluation of the conversion factors 
between ADC units and energy deposits, ensuring linearity over each 
gain range (TASC has four gain ranges for each channel), 
and provides a seamless transition between neighboring gain ranges. 
Temporal gain variations occurring during long time observations are also corrected for in the calibration procedure~\cite{CALET2017}. 
The errors at each calibration step, such as the correction of position and temperature 
dependence, consistency between energy deposit peaks of noninteracting protons and helium, linear fit error of each gain range, and 
gain ratio measurements, as well as slope extrapolation, are included 
in the estimation of the energy resolution.
As a result, a very high resolution of 2\% or better is achieved above 20~GeV~\cite{asaoka2017}.
It should be noted that even with such a detailed calibration, the determining factor for the energy
resolution is the calibration uncertainty, as the intrinsic resolution of CALET is $\sim$1\% as for DAMPE~\cite{DAMPE-Mission}. 
Intrinsic resolution refers to the detector's capability by design, taking advantage of
the thick, fully-active total absorption calorimeter.
Also important is the fact that the calibration error in the lower gain ranges is crucial 
for the spectrum measurements in the TeV range.

We use the ``electromagnetic shower tracking'' algorithm~\cite{akaike2013} to 
reconstruct the shower axis of each event, taking advantage of 
the electromagnetic shower shape and IMC design concept. 
As input for the electron identification, 
well-reconstructed and well-contained single-charged events 
are preselected by (1) an off-line trigger confirmation,  (2)  a geometrical condition, 
(3) a track quality cut to ensure reconstruction accuracy, (4) a charge selection using CHD, (5) a longitudinal  shower development  and  (6) a lateral shower containment consistent with those expected for electromagnetic cascades.  
The geometrical condition in our analysis is divided into four categories
(A, B, C, D), 
depending on which detector components are penetrated by the shower axis,
explained in detail in Fig.~1 of
the Supplemental Material~\cite{CALET2018-SM-arXiv} and its caption. 
In brief, A+B are fully contained events, while category C adds events incident from the IMC sides,
and D adds events exiting through the sides of TASC. 
For events not crossing the CHD, we use the energy deposit of the first hit IMC layer to determine their charge.

The energy of incident electrons is reconstructed using the energy correction function,
which converts the energy deposit information of TASC and IMC into primary energy for each geometrical condition. 
In order to identify electrons and to study systematic uncertainties in the 
electron identification, we applied two methods: 
a simple two parameter cut and a multivariate analysis
based on boosted decision trees (BDTs).
The details concerning these methods are explained in the Supplemental Material of Ref.~\cite{CALET2017}. 

\begin{figure}[bt!] 
\begin{center}
\includegraphics[width=\hsize]{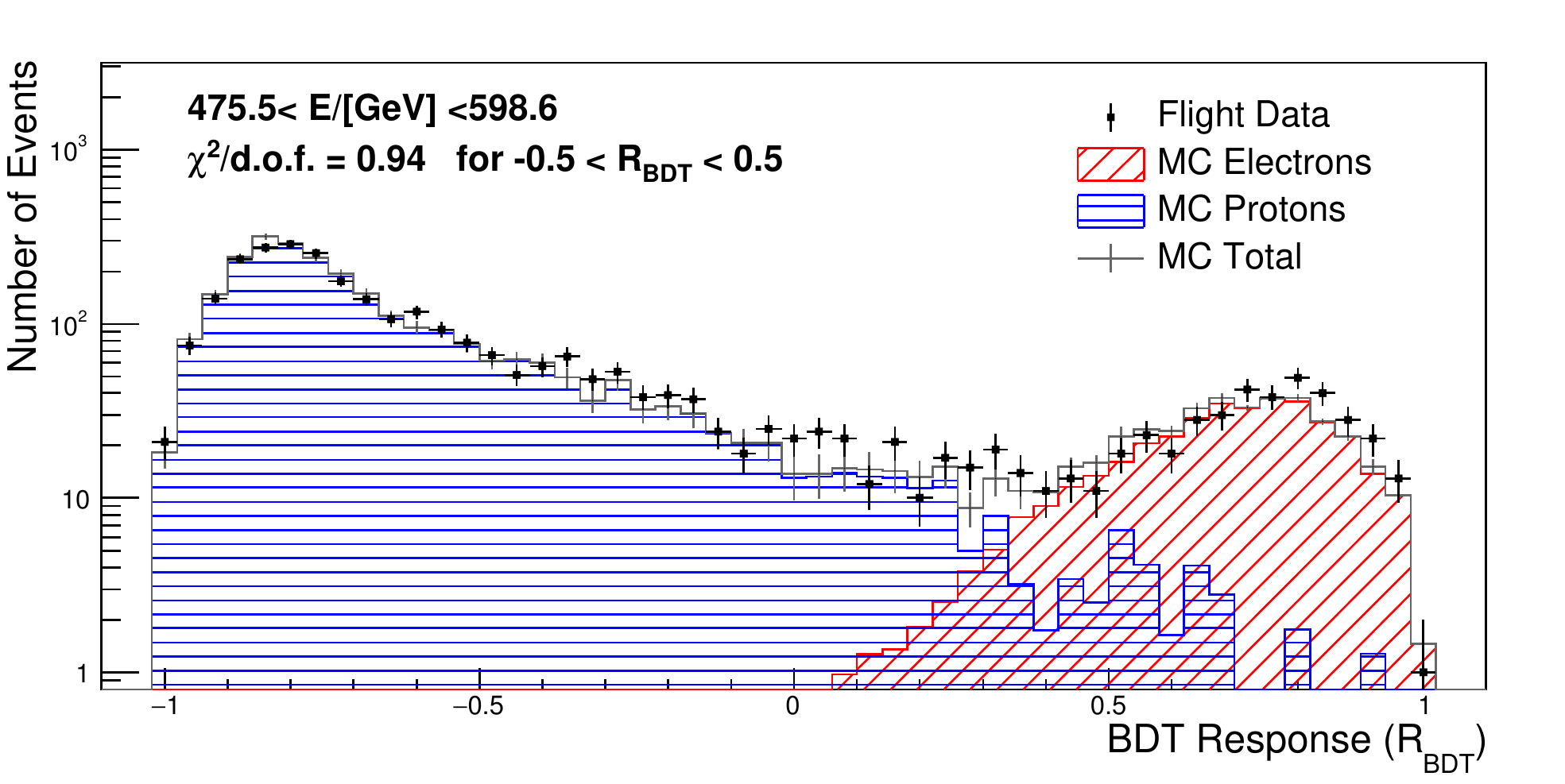}
\caption{An example of BDT response distributions in the 476 $< E < $ 599~GeV bin including all acceptance conditions A, B, C and D.
The BDT response distributions for the TeV region are shown in Fig.~2 of the Supplemental Material~\cite{CALET2018-SM-arXiv}.
}
\label{fig:distBDT}
\end{center}
\end{figure}

Calculation of event selection efficiencies, BDT training, and 
estimation of proton background contamination 
are carried out separately for each geometrical condition,
and combined in the end to obtain the final spectrum.
Considering the fact that the lower energy region is dominated by systematics in our analysis and therefore 
more statistics would not significantly improve the precision of our data,  
the acceptance conditions C and D are only included in 
the higher energy region above 475~GeV.
An example of a BDT response distribution including all acceptance conditions is shown in Fig.~\ref{fig:distBDT}. 
In the final electron sample, the resultant contamination ratios of protons are  $\sim$5\% 
up to 1~TeV, and 10\%--20\% in the 1--4.8~TeV region, while keeping a constant high efficiency of 80\% for electrons.
The number of electron candidates in the highest energy bin is seven.

The absolute energy scale was calibrated and shifted by $+$3.5\%~\cite{CALET2017} 
as a result of a study of the geomagnetic cutoff energy~\cite{Fermi2012}. 
Since the full dynamic range calibration~\cite{asaoka2017} was carried out with a scale-free method, 
its validity holds regardless of the absolute scale uncertainty.

\section{Systematic Uncertainties}
As discussed in detail in Ref.~\cite{CALET2017} and its Supplemental Material,
systematic uncertainties in our flux measurements can be divided into three categories, i.e.,
energy scale uncertainty, absolute normalization, and energy dependent uncertainties.
As per the energy dependent systematics, we have identified  
the following contributions: trigger efficiency (below 30~GeV), 
BDT stability, tracking, charge identification, electron identification, 
and MC model dependence.

BDT stability is evaluated from the stability of the resultant flux 
for 100 independent training samples and 
for BDT cut efficiency variation from 70\% to 90\% 
in 1\% steps for each corresponding test sample.
Upper and lower panels of Fig.~2
in the Supplemental Material~\cite{CALET2018-SM-arXiv} show an example for the  
stability of the BDT analysis in the $949 < E < 1194$~GeV bin and its energy dependence, respectively,
where good stability over a wide range of 
efficiency factors and number of training samples is demonstrated.
Dependence on tracking, charge identification, electron identification
and MC model is estimated by using the difference of the resultant flux 
between representative algorithms or methods, i.e., electromagnetic shower tracking
vs combinatorial Kalman filter tracking~\cite{paolo2017} algorithms, CHD vs IMC charge identification methods,
simple two parameter cut vs BDT cut, and the use of EPICS vs Geant4, respectively.
The obtained energy dependence of the relative flux difference in each case 
is fitted with a suitable log-polynomial function to mitigate 
statistical fluctuations, as shown in Fig.~3 of the Supplemental Material~\cite{CALET2018-SM-arXiv}. 
Systematic effects up to a few percent are seen in the energy range below the TeV region.
Statistical fluctuations are the most important limiting factor for estimating systematic errors in the 
TeV region, as indicated by the changes in the energy dependence of the MC model comparison
from the previous publication~\cite{CALET2017}. 
By adding a factor of two more statistics in the highest energy region, 
the deviation in the 2--3~TeV bin changed significantly from the previous estimate, 
though by a smaller extent than the statistical error on the flux.

Since other selections, such as the track quality cut and shower concentration cuts, did not have a significant energy dependence, they were treated as uncertainties in the absolute normalization. 
Their contribution to the uncertainty in the absolute normalization was determined to be a very small part of the total. 
The total uncertainty in the absolute normalization was estimated to be 3.2\%. 
A detailed breakdown of this uncertainty is given 
in the Supplemental Material of Ref.~\cite{CALET2017}.
The high-energy trigger efficiency was verified by using data obtained with the
low-energy trigger (1~GeV threshold) in the low rigidity cutoff region 
below 6~GV. By comparing the flux with and without off-line trigger confirmation,
the systematic uncertainty from trigger efficiency is estimated to be 2.4\% below 30~GeV,
mainly limited by the available low-energy triggered data, 
and is negligible above this energy. 
The resultant flux for each of the acceptance conditions used in this analysis 
is consistent within the statistical uncertainty,
indicating that there are no significant systematic deviations among the acceptance conditions.

\section{Electron + Positron Spectrum}
\begin{figure*}[bth!]
\begin{center}
\includegraphics[width=\hsize]{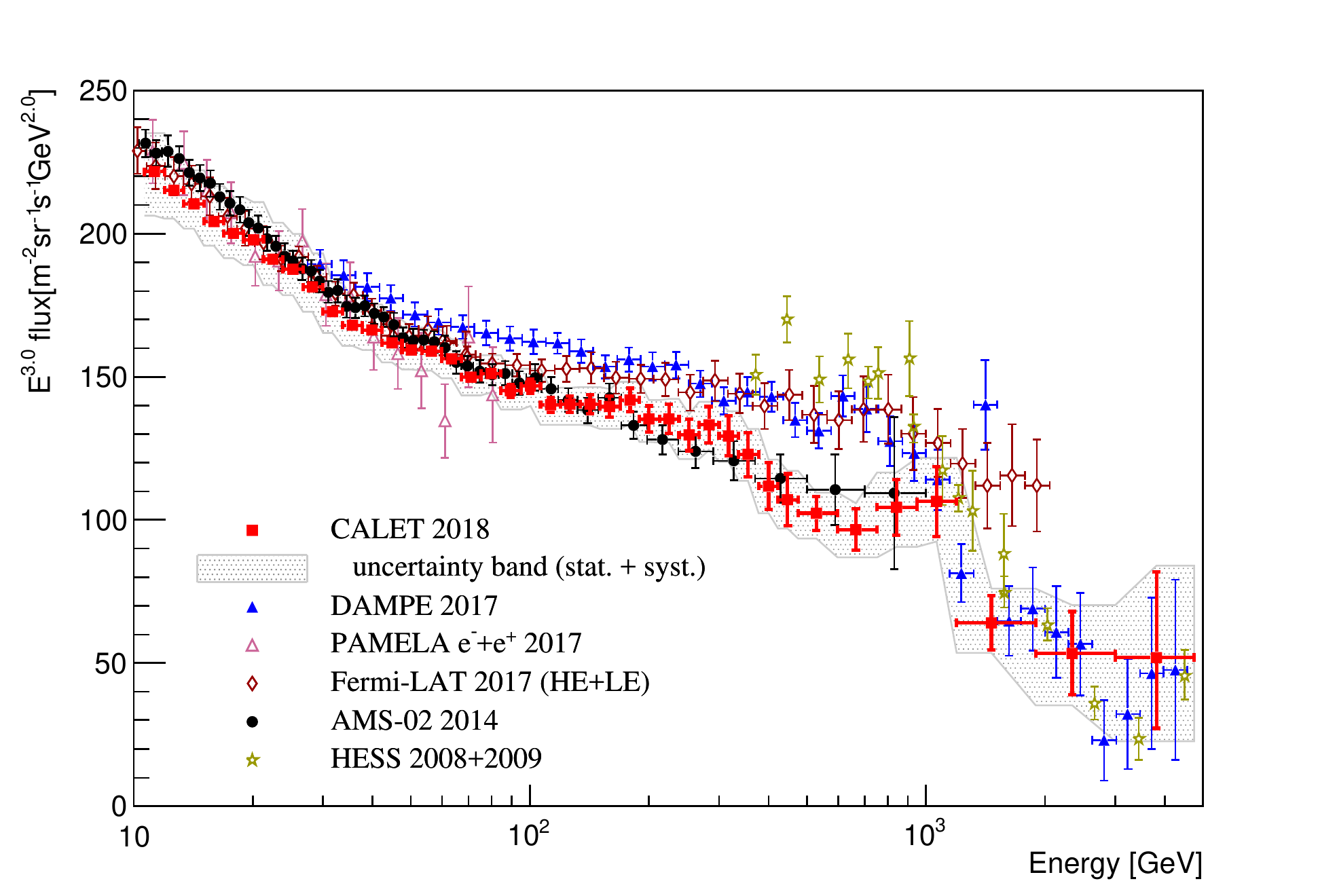}
\caption{Cosmic-ray all-electron spectrum measured by CALET from 10.6~GeV to 4.75~TeV using the same energy binning as in our previous publication~\cite{CALET2017}, 
where the gray band indicates the quadratic sum of statistical and systematic errors (not including the uncertainty on the energy scale). 
Also plotted are direct measurements in space~\cite{DAMPE2017,Pamela-e,Fermi2017-e,AMS02-e} and from ground-based experiments~\cite{HESS2008,HESS2009}. 
} 
\label{fig:binCALET}
\end{center}
\end{figure*}
Figure~\ref{fig:binCALET} shows the 
extended 
electron and positron spectrum obtained with CALET 
using the same energy binning as in our previous publication, except for adding one extra bin at the
high-energy end.
The error bars along horizontal and vertical axes indicate bin width 
and statistical errors, respectively. 
The gray band is representative of the quadratic sum of statistical and systematic errors, 
using the same definition as the one given in Ref.~\cite{CALET2017}. 
Systematic errors include
errors in the absolute normalization and energy dependent ones, except for the energy scale uncertainty. 
The energy dependent errors include those obtained from BDT stability, trigger efficiency in the low-energy region, 
tracking dependence, dependence on charge and electron identification methods and MC model dependence.
In more refined interpretation studies, the latter four contributions could be treated as nuisance parameters 
while the first two components must be added in 
quadrature to the statistical errors. 
Conservatively, all of them are included in the total error estimate
in Fig.~\ref{fig:binCALET}. 
The measured all-electron flux including statistical errors 
and a detailed breakdown of the systematic errors into their components
is tabulated in Table~1 of the Supplemental Material~\cite{CALET2018-SM-arXiv}.

Comparing with other recent experiments [AMS-02, Fermi Large Area Telescope (Fermi-LAT) and DAMPE], 
our spectrum shows good agreement with AMS-02 data below 1~TeV.
In the energy region from 40 to 300~GeV,
the power-law index of CALET's spectrum is found to be 
$-3.12 \pm 0.02$, which is consistent with other experiments within errors.
However, the spectrum is considerably softer from 300 to 600~GeV than
the spectra measured by DAMPE and Fermi-LAT.
The CALET 
results exhibit a lower flux than those of DAMPE and Fermi-LAT from 
300~GeV up to near 1~TeV.
In this
region, 
a difference 
is noticeable between two groups of
measurements with internal consistency 
within 
each group: CALET and AMS-02
vs Fermi-LAT and DAMPE, 
indicating the presence of
unknown systematic effects.

\begin{figure}[bth!]
\begin{center}
\includegraphics[width=\hsize]{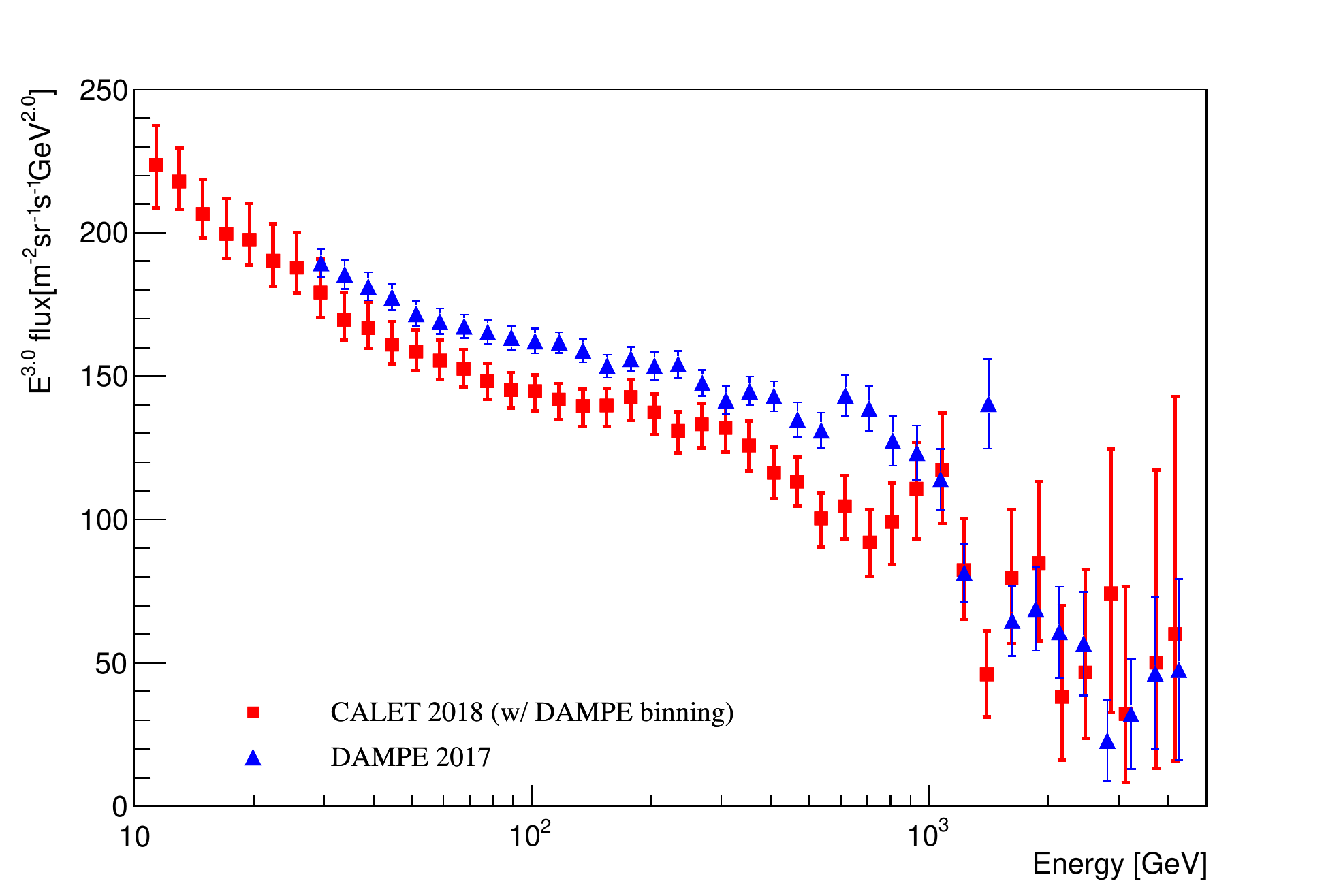}
\caption{
Cosmic-ray all-electron spectrum measured by CALET from 10.6~GeV to 4.57~TeV 
using the same energy binning as DAMPE's result~\cite{DAMPE2017} and compared with it. 
The error bars indicate the quadratic sum of 
statistical and systematic errors (not including the uncertainty on the energy scale). 
}
\label{fig:binDAMPE}
\end{center}
\end{figure}
In Fig.~\ref{fig:binDAMPE} 
we have adopted exactly the same energy binning as DAMPE to show our spectrum.
The tabulated flux for this energy binning with 
a
detailed breakdown of systematics 
is also shown in Table~2 of the Supplemental Material~\cite{CALET2018-SM-arXiv}. 
To check if the CALET spectrum is consistent 
with a possible break at 0.9 TeV, as suggested by DAMPE's observations,
we fit our spectrum with a smoothly broken power-law model~\cite{DAMPE2017}
in the energy range 
from 
55~GeV to 2.63~TeV,
while fixing the break energy at 914~GeV. 
A broken 
power law steepening from $-3.15 \pm 0.02$ to $-3.81 \pm 0.32$ fits 
our data well, with $\chi^2 = 17.0$ and number of degrees of freedom 
(NDF) equal to 25; this result is consistent with DAMPE 
regarding the spectral index change of $0.7 \pm 0.3$. 
However, a single power-law fit over the same energy range gives an 
index $-3.17 \pm 0.02$ with $\chi^2$/NDF = 26.5/26, not a significantly 
poorer goodness of fit than obtained with the broken power law.
The fitting results are shown in Fig. 5 of the 
Supplemental Material~\cite{CALET2018-SM-arXiv}, including a fit with an 
exponentially cutoff power law~\cite{Fermi2017-e}. 

On the other hand, 
the flux in the 1.4 TeV bin of DAMPE's spectrum, 
which might imply a peak structure, 
is not compatible with CALET results at a level of 4~$\sigma$ significance,
including the systematic errors from both experiments.
Since a sharp peak 
in a single bin 
could be an artifact 
due to binning effects,
we have studied this kind of effect
as shown in Fig.~6 of the Supplemental Material~\cite{CALET2018-SM-arXiv}
and explained in its caption.
The result of this study excludes 
with good significance the hypothesis of the presence of a 
peaklike structure in our data. 
Furthermore, bin-to-bin migration and related effects are found to be 
negligible 
when compared with
our estimated systematic uncertainties.


In conclusion, we 
extended our previous result~\cite{CALET2017} on 
the
CALET all-electron spectrum both in energy (to 4.8 TeV) and in acceptance, 
with an approximate increase by a factor of 2 of the statistics in the higher energy region. 
The data in the TeV region show a suppression of the flux compatible with the DAMPE results. 
However, the accuracy of the break's sharpness and position, and of the spectral shape above 1~TeV,  will improve by better statistics and a further reduction of the systematic errors based on the analysis of additional flight data during the ongoing five-year (or more) observation. 
By specifying the breakdown of systematic uncertainties,
our extended all-electron spectrum together with 
the 
AMS-02
positron flux measurement~\cite{AMS02-ep}
provides
essential information
to investigate spectral features in the framework of pulsars and/or dark matter inspired models. 

\section{Acknowledgments}
\begin{acknowledgments}
We gratefully acknowledge JAXA's contributions to the development of CALET and to the
operations onboard the International Space Station. 
We also wish to express our sincere gratitude to ASI and NASA for
their support of the CALET project. This work was supported in part by a JSPS Grant-in-Aid for
Scientific Research (S) (No. 26220708) and by the MEXT-Supported Program for the Strategic
Research Foundation at Private Universities (2011-2015) (No. S1101021) at Waseda University.
The CALET effort in the United States is supported by NASA through Grants No. NNX16AB99G, No. NNX16AC02G, and No. NNH14ZDA001N-APRA-0075.
\end{acknowledgments}

\providecommand{\noopsort}[1]{}\providecommand{\singleletter}[1]{#1}%
%
\widetext
\clearpage
\begin{center}
\end{center}
\setcounter{equation}{0}
\setcounter{figure}{0}
\setcounter{table}{0}
\setcounter{page}{1}
\makeatletter
\renewcommand{\theequation}{S\arabic{equation}}
\renewcommand{\thefigure}{S\arabic{figure}}
\renewcommand{\bibnumfmt}[1]{[S#1]}
\renewcommand{\citenumfont}[1]{S#1}

\begin{center}
\textbf{\Large Extended Measurements of Cosmic-ray Electron and Positron Spectrum \\
from 11 GeV to 4.8 TeV with the Calorimetric Electron Telescope \\
on the International Space Station\\
Supplemental online material.}

\end{center}
\vspace*{0.5cm}
Supplemental material concerning ``Extended Measurements of Cosmic-ray Electron and Positron Spectrum from 11 GeV to 4.8 TeV with the Calorimetric Electron Telescope on the International Space Station.''
\vspace*{1cm}

\begin{figure}[hbt!]
\begin{center}
\includegraphics[bb=0 0 620 200, width=1.00\linewidth]{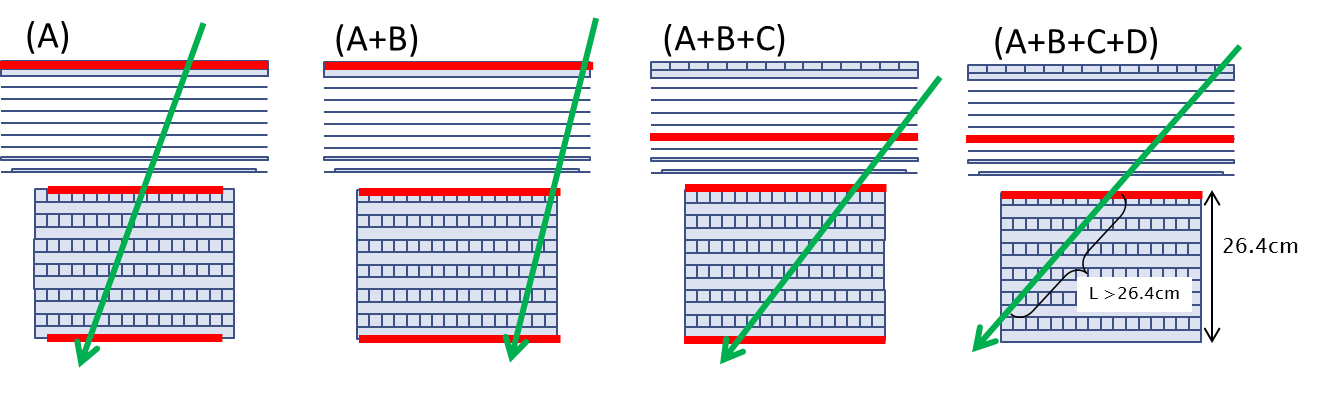}
\caption{
Classification of four geometrical condition types. The red bars indicate the geometrical requirements 
for the particle trajectory at different detector layers. 
In each of the plots, the green arrow shows an example of a shower axis which is allowed by the extention of the geometrical condition. 
Geometrical condition A requires the shower axis
to cross the TASC top and bottom layers except for a 2~cm outer margin, while in geometrical condition A$+$B 
the shower axis is allowed to cross the edge of TASC. 
In geometrical condition A$+$B$+$C, incidence from the side is allowed, 
but passage through the IMC 5th layer is required. 
Geometrical condition A$+$B$+$C$+$D does not require crossing of the TASC bottom layer, but it is required that 
the track length in TASC is greater than 26.4~cm. 
Each of A, B, C and D are defined exclusively, e.g. an event fulfilling condition A is not listed as 
fulfilling condition B as well.
}
\label{fig:geom}
\end{center}
\end{figure}

\begin{figure}[bt!] 
\begin{center}
\includegraphics[width=\hsize]{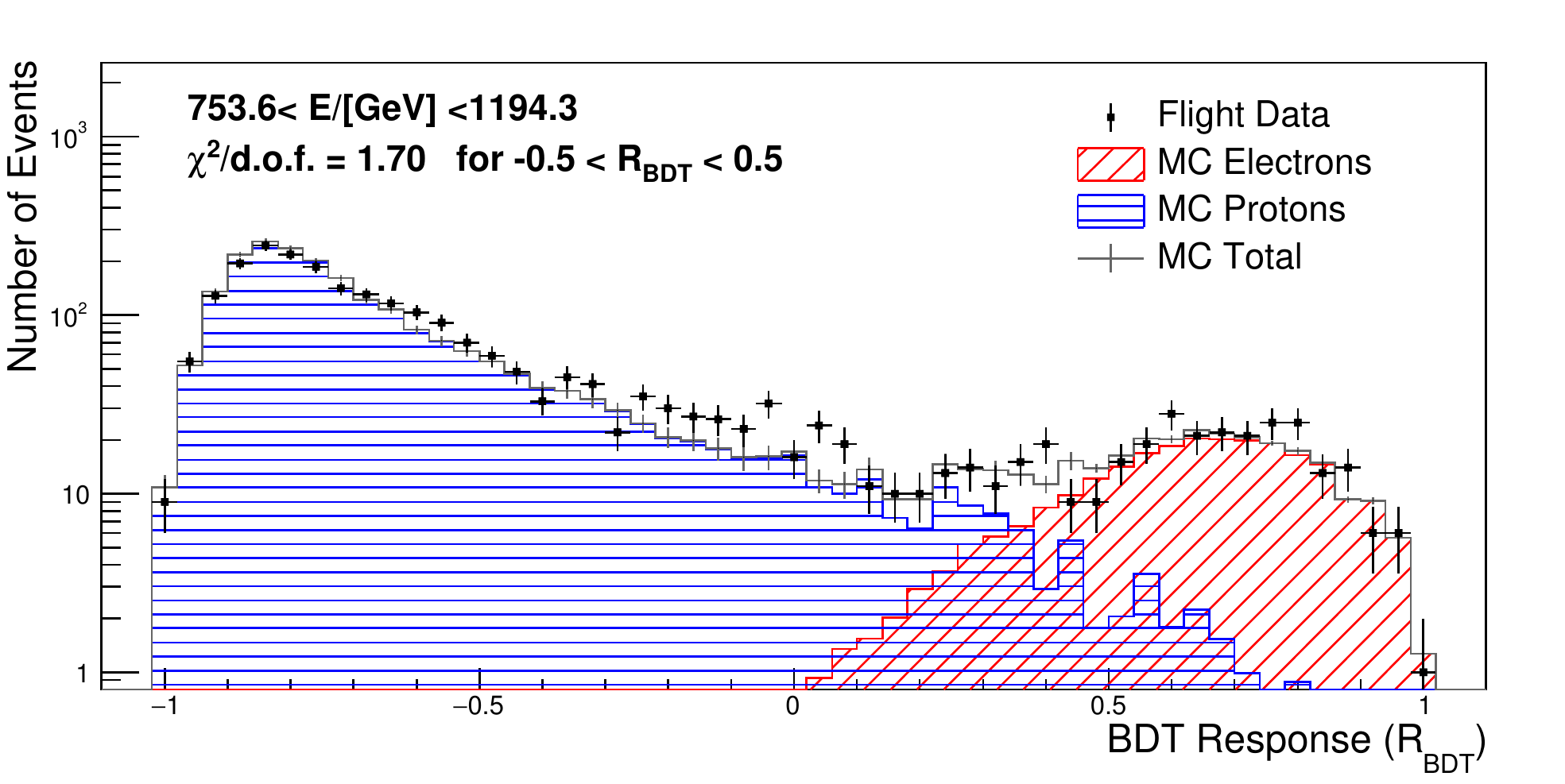}
\includegraphics[width=\hsize]{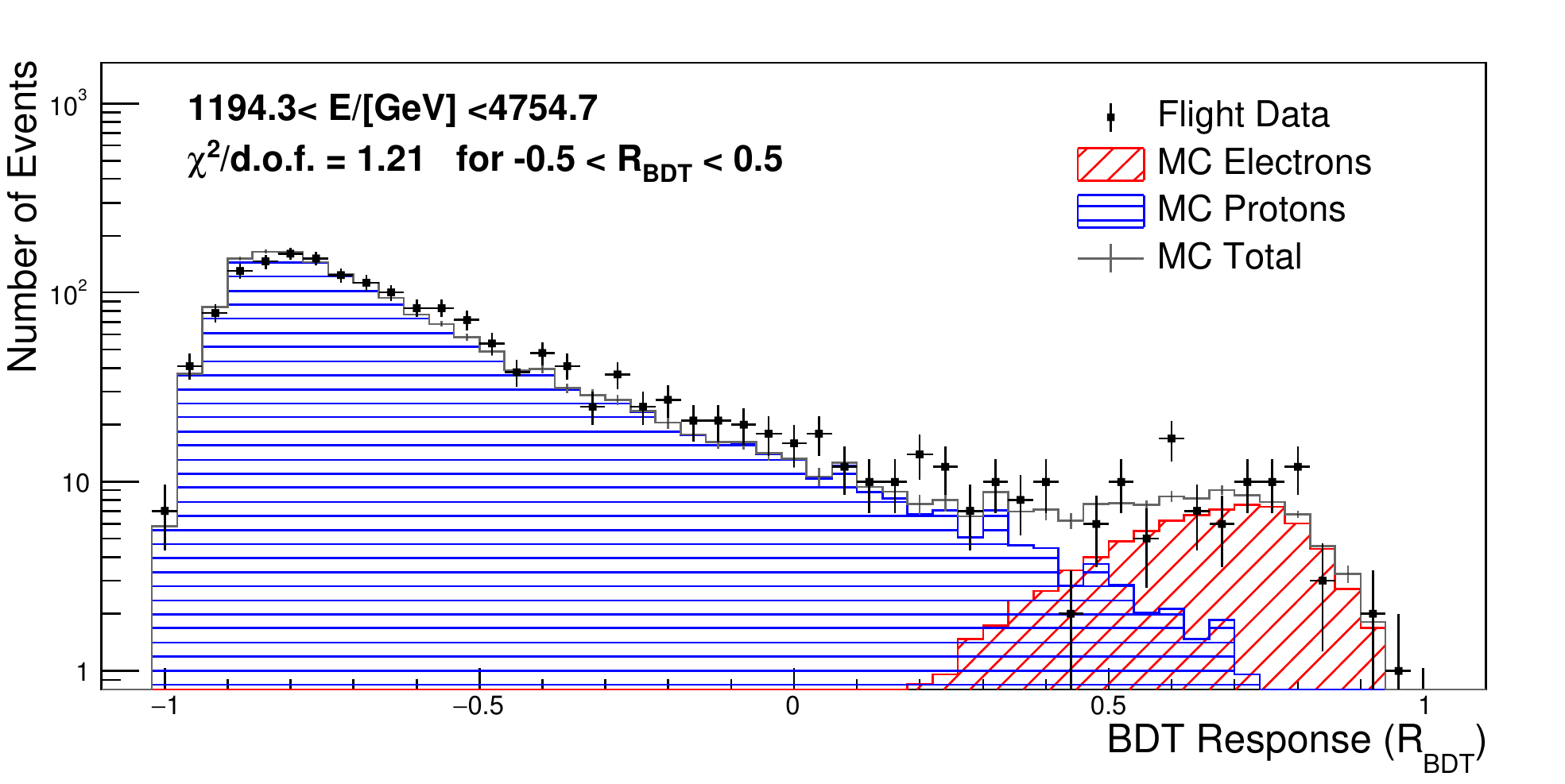}
\caption{Examples of BDT response ($R_{\rm BDT}$) distributions in the 754 $< E < $ 1194~GeV ({\it Top}) 
and 1194 $< E < $ 4575~GeV ({\it Bottom}) bins including all acceptance conditions A, B, C and D.
While there is 
noticeable 
discrepancies in the -0.3 $< R_{\rm BDT} <$ 0 region, 
their possible effects 
to the resultant spectrum are included in the systematic uncertainty concerning the BDT stability.
}
\label{fig:SM_distBDT}
\end{center}
\end{figure}

\begin{figure}[hbt] 
\begin{center}
\includegraphics[width=0.9\hsize]{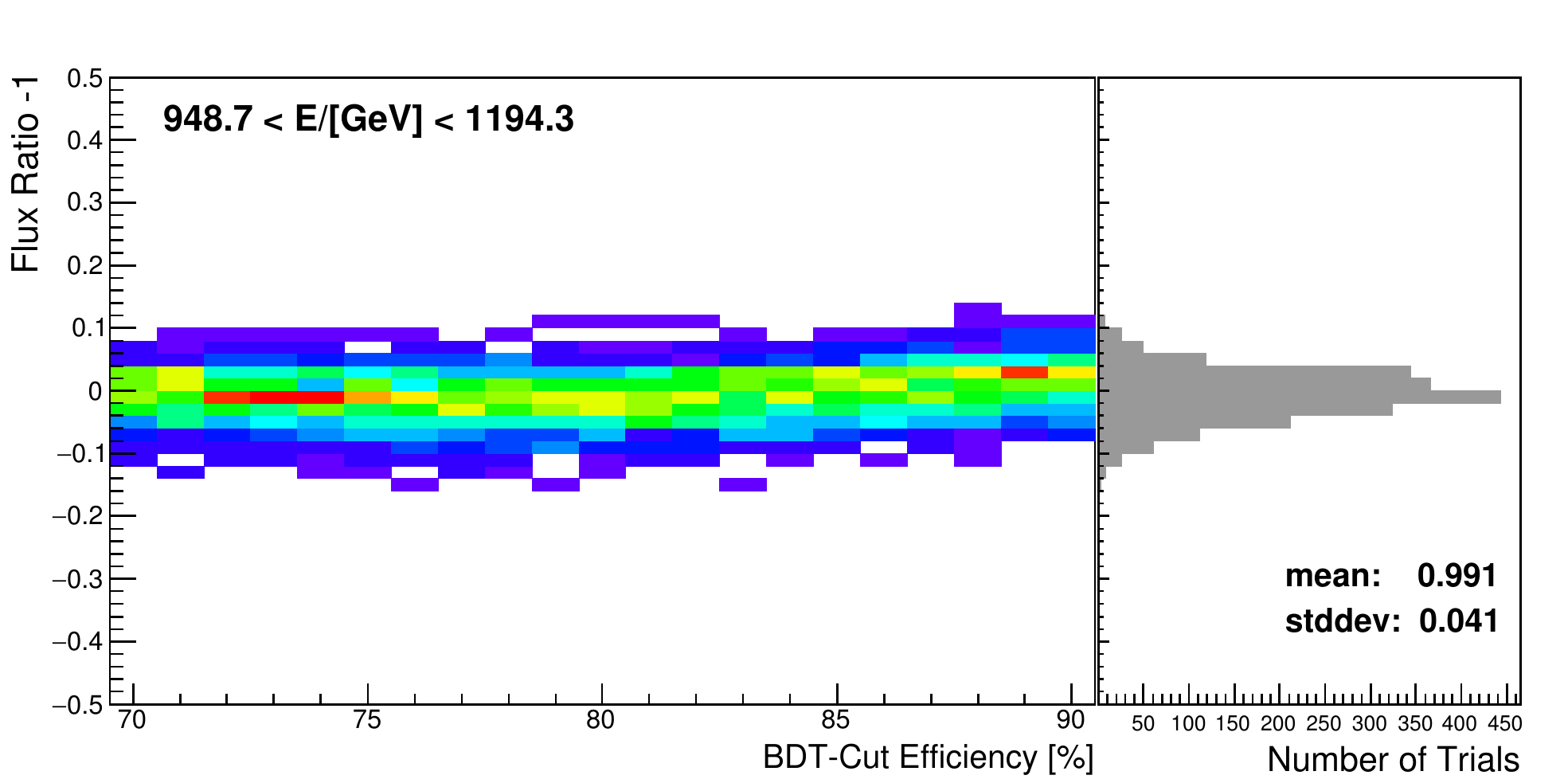}
\includegraphics[width=\hsize]{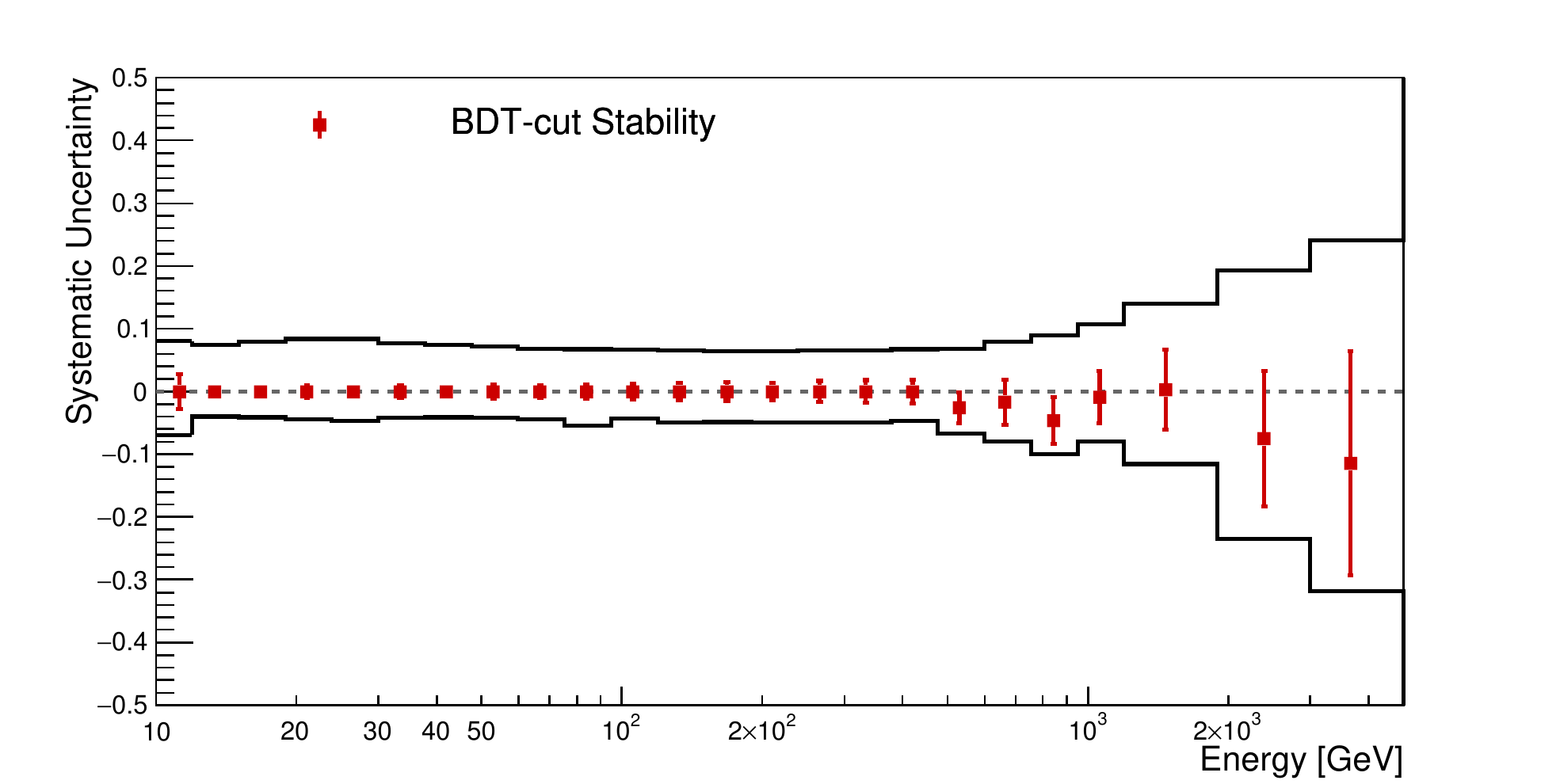}
\caption{
({\it Top}) Stability of BDT analysis with respect to independent training samples and BDT-cut efficiency 
in the $949<E<1194$~GeV bin. Color maps show
the flux ratio dependence on efficiency, where the bin value (number of trials) 
increases as color changes from violet, blue, green, yellow to red.
A projection onto the $Y$-axis is shown as a rotated histogram (in gray color). 
({\it Bottom})
Energy dependence of systematic uncertainties. The red squares represent the systematic uncertainties stemming from the electron identification based on BDT.  The bands defined by black lines show the sum in quadrature of all the sources of systematics, except the energy scale uncertainties.
}
\label{fig:systBDT}
\end{center}
\end{figure}
\begin{figure}[hbt] 
\begin{center}
\includegraphics[width=\hsize]{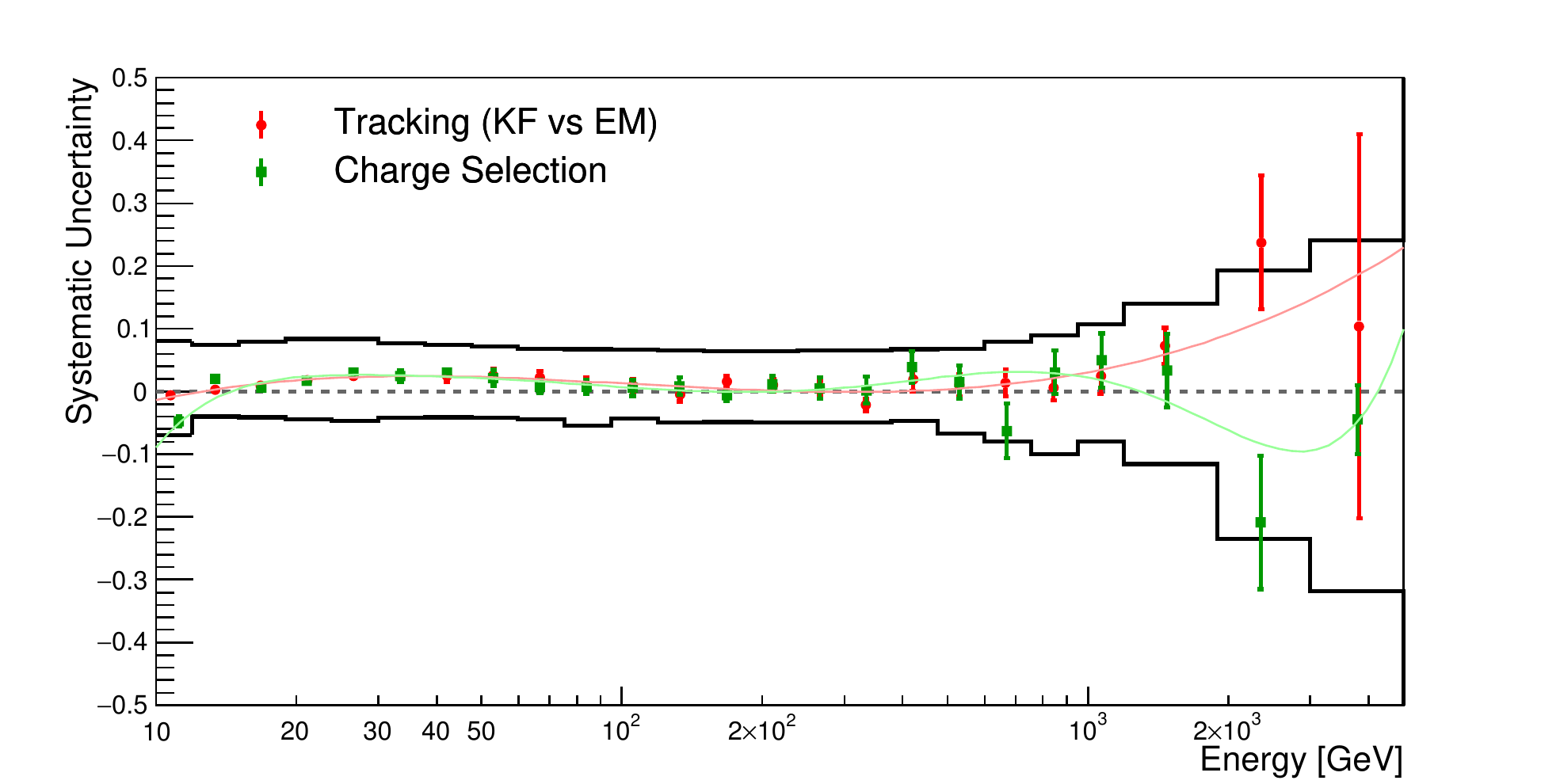}
\includegraphics[width=\hsize]{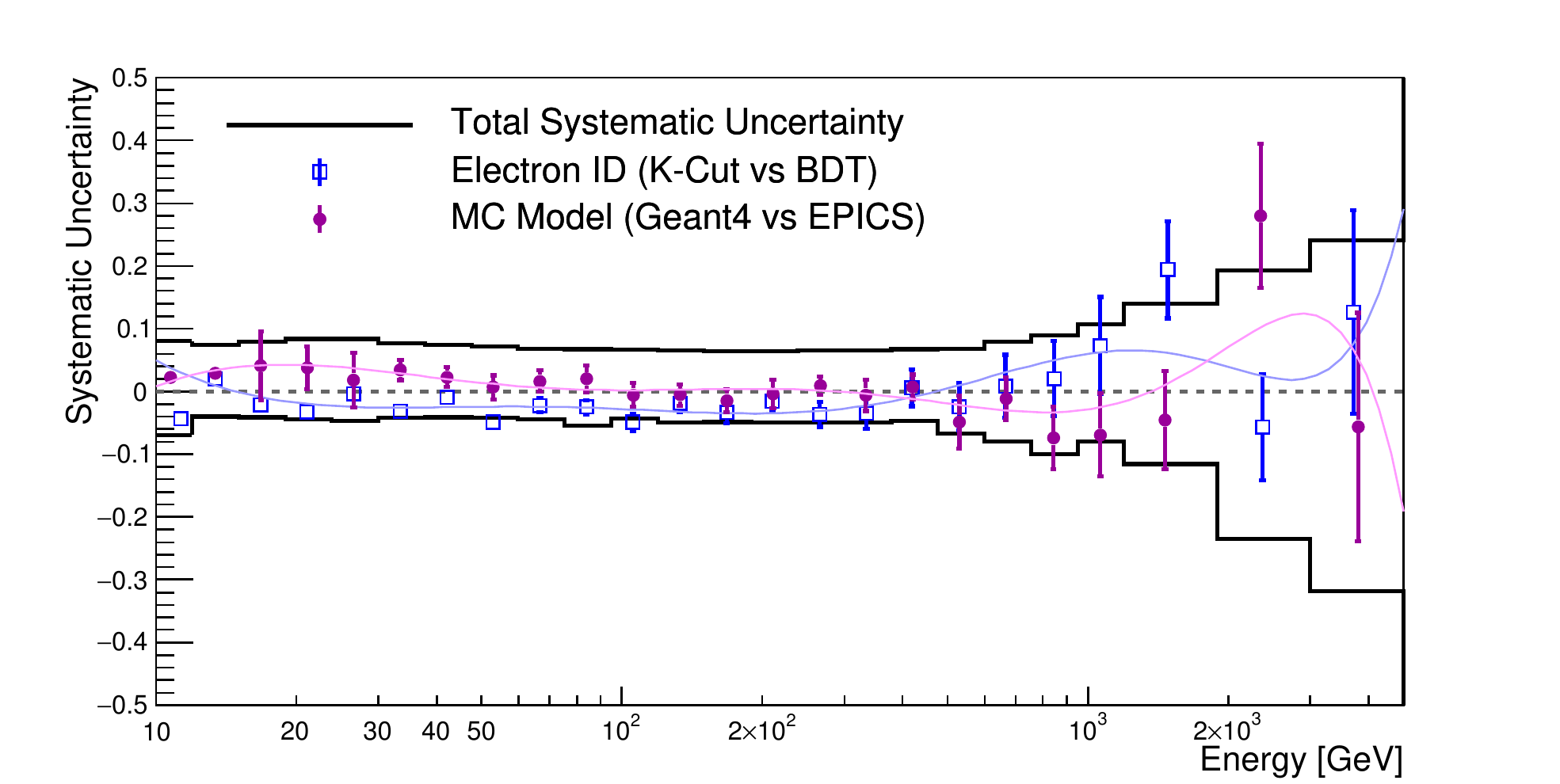}
\caption{
({\it Top}) Energy dependence of systematic uncertainties in tracking algorithms
(Electromagnetic shower tracking vs combinatorial Kalman filter tracking) 
and charge identification methods (CHD vs IMC),
({\it Bottom}) Energy dependence of systematic uncertainties in electron identification methods 
(K-estimator vs BDT) and MC models (Geant4 vs EPICS).
The data points are fitted with log-polynomial functions to mitigate the effect of statistical 
fluctuations while preserving possible energy dependent structures.
Fit functions are shown as curves and are used to estimate
energy dependent systematic uncertainties.
}
\label{fig:systEDep}
\end{center}
\end{figure}
\clearpage

\renewcommand{\arraystretch}{1.25}
\begin{table*}
\caption{Table of CALET electron plus positron spectrum. 
Mean energy is calculated using the candidate events in the energy bin.
For the flux the first and second 
errors represent the statistical uncertainties (68\% confidence level) and
systematic uncertainties, respectively, while
uncertainty in the energy scale is not included.
Detailed breakdown of systematic errors is included where
$\sigma_{\rm BDT}$, 
$\sigma_{\rm trig.}$, 
$\Delta_{\rm norm.}$, 
$\Delta_{\rm trk.}$, 
$\Delta_{\rm chg.}$, 
$\Delta_{\rm ID}$ and
$\Delta_{\rm MC}$ denote systematic errors due to
BDT stability, trigger, absolute normalization, tracking, charge identification,
electron identification, and MC model dependence, respectively.
While the first two components must be added in quadrature to the statistical errors in the spectral analysis, 
the latter five contributions could be treated as nuisance parameters, 
by introducing correction factors corresponding to each component.
Energy dependence (or non-dependence) of each contribution is already determined, 
constraining the possible corrections due to the nuisance parameters. 
Although $\Delta_{\rm norm.}$ can be ignored in a spectral study only using CALET data, it should also be treated as a nuisance parameter in a combined analysis with the positron spectrum, for example. 
\label{tab:ele}}
\begin{ruledtabular}
\begin{tabular}{cccrrrrrrr}
Energy Bin & Mean Energy & Flux & \multicolumn{7}{c}{Systematic Uncertainties (relative to flux)} \\
(GeV)  & (GeV) & (m$^{-2}$sr$^{-1}$s$^{-1}$GeV$^{-1}$) & $\sigma_{\rm BDT}$ & $\sigma_{\rm trig.}$ & $\Delta_{\rm norm.}$  & $\Delta_{\rm trk.}$ & $\Delta_{\rm chg.}$ & $\Delta_{\rm ID}$ & $\Delta_{\rm MC}$ \\ 
\hline
10.6--11.9 & $11.3$ &$(1.543 \, \pm 0.004 \, _{-0.106}^{+0.094}) \times 10^{-1}$ & 0.028 & 0.024 & 0.032 & -0.006 & -0.048 & 0.031 & 0.021\\
11.9--13.4 & $12.6$ &$(1.065 \, \pm 0.003 \, _{-0.049}^{+0.057}) \times 10^{-1}$ & 0.010 & 0.024 & 0.032 & 0.000 & -0.021 & 0.016 & 0.030\\
13.4--15.0 & $14.2$ &$(7.388 \, \pm 0.023 \, _{-0.305}^{+0.410}) \times 10^{-2}$ & 0.010 & 0.024 & 0.032 & 0.006 & -0.003 & 0.004 & 0.037\\
15.0--16.9 & $15.9$ &$(5.073 \, \pm 0.018 \, _{-0.211}^{+0.302}) \times 10^{-2}$ & 0.010 & 0.024 & 0.032 & 0.011 & 0.010 & -0.005 & 0.041\\
16.9--18.9 & $17.8$ &$(3.521 \, \pm 0.014 \, _{-0.152}^{+0.223}) \times 10^{-2}$ & 0.010 & 0.024 & 0.032 & 0.014 & 0.018 & -0.013 & 0.043\\
18.9--21.2 & $20.0$ &$(2.468 \, \pm 0.011 \, _{-0.111}^{+0.162}) \times 10^{-2}$ & 0.010 & 0.024 & 0.032 & 0.018 & 0.022 & -0.018 & 0.043\\
21.2--23.8 & $22.5$ &$(1.687 \, \pm 0.008 \, _{-0.079}^{+0.112}) \times 10^{-2}$ & 0.010 & 0.024 & 0.032 & 0.020 & 0.025 & -0.022 & 0.041\\
23.8--26.7 & $25.2$ &$(1.171 \, \pm 0.006 \, _{-0.055}^{+0.077}) \times 10^{-2}$ & 0.008 & 0.024 & 0.032 & 0.022 & 0.026 & -0.024 & 0.038\\
26.7--30.0 & $28.3$ &$(8.029 \, \pm 0.034 \, _{-0.385}^{+0.516}) \times 10^{-3}$ & 0.008 & 0.024 & 0.032 & 0.023 & 0.026 & -0.025 & 0.035\\
30.0--33.7 & $31.7$ &$(5.413 \, \pm 0.026 \, _{-0.229}^{+0.313}) \times 10^{-3}$ & 0.011 & 0.000 & 0.032 & 0.024 & 0.026 & -0.026 & 0.031\\
33.7--37.8 & $35.6$ &$(3.721 \, \pm 0.020 \, _{-0.157}^{+0.206}) \times 10^{-3}$ & 0.011 & 0.000 & 0.032 & 0.025 & 0.025 & -0.026 & 0.026\\
37.8--42.4 & $39.9$ &$(2.612 \, \pm 0.016 \, _{-0.109}^{+0.137}) \times 10^{-3}$ & 0.010 & 0.000 & 0.032 & 0.024 & 0.024 & -0.025 & 0.022\\
42.4--47.5 & $44.8$ &$(1.798 \, \pm 0.013 \, _{-0.075}^{+0.090}) \times 10^{-3}$ & 0.010 & 0.000 & 0.032 & 0.024 & 0.023 & -0.025 & 0.018\\
47.5--53.3 & $50.3$ &$(1.255 \, \pm 0.010 \, _{-0.052}^{+0.060}) \times 10^{-3}$ & 0.011 & 0.000 & 0.032 & 0.023 & 0.021 & -0.024 & 0.014\\
53.3--59.9 & $56.4$ &$(8.863 \, \pm 0.078 \, _{-0.367}^{+0.404}) \times 10^{-4}$ & 0.011 & 0.000 & 0.032 & 0.022 & 0.019 & -0.024 & 0.010\\
59.9--67.2 & $63.3$ &$(6.157 \, \pm 0.061 \, _{-0.254}^{+0.266}) \times 10^{-4}$ & 0.010 & 0.000 & 0.032 & 0.020 & 0.017 & -0.024 & 0.007\\
67.2--75.4 & $71.0$ &$(4.188 \, \pm 0.048 \, _{-0.174}^{+0.173}) \times 10^{-4}$ & 0.010 & 0.000 & 0.032 & 0.019 & 0.015 & -0.025 & 0.005\\
75.4--84.6 & $79.7$ &$(2.984 \, \pm 0.038 \, _{-0.126}^{+0.119}) \times 10^{-4}$ & 0.012 & 0.000 & 0.032 & 0.017 & 0.012 & -0.025 & 0.003\\
84.6--94.9 & $89.4$ &$(2.032 \, \pm 0.030 \, _{-0.087}^{+0.078}) \times 10^{-4}$ & 0.012 & 0.000 & 0.032 & 0.015 & 0.010 & -0.027 & 0.002\\
94.9--106.4 & $100.4$ &$(1.45 \, \pm 0.02 \, _{-0.06}^{+0.05}) \times 10^{-4}$ & 0.013 & 0.000 & 0.032 & 0.013 & 0.007 & -0.028 & 0.002\\
106.4--119.4 & $112.6$ &$(9.82 \, \pm 0.18 \, _{-0.45}^{+0.36}) \times 10^{-5}$ & 0.013 & 0.000 & 0.032 & 0.011 & 0.005 & -0.030 & 0.002\\
119.4--134.0 & $126.2$ &$(6.98 \, \pm 0.15 \, _{-0.33}^{+0.25}) \times 10^{-5}$ & 0.014 & 0.000 & 0.032 & 0.009 & 0.003 & -0.031 & 0.003\\
134.0--150.4 & $141.7$ &$(4.93 \, \pm 0.12 \, _{-0.24}^{+0.18}) \times 10^{-5}$ & 0.014 & 0.000 & 0.032 & 0.007 & 0.001 & -0.033 & 0.003\\
150.4--168.7 & $159.0$ &$(3.47 \, \pm 0.09 \, _{-0.17}^{+0.12}) \times 10^{-5}$ & 0.014 & 0.000 & 0.032 & 0.005 & -0.000 & -0.034 & 0.004\\
168.7--189.3 & $178.8$ &$(2.48 \, \pm 0.07 \, _{-0.12}^{+0.09}) \times 10^{-5}$ & 0.014 & 0.000 & 0.032 & 0.003 & -0.001 & -0.035 & 0.004\\
189.3--212.4 & $200.1$ &$(1.69 \, \pm 0.06 \, _{-0.08}^{+0.06}) \times 10^{-5}$ & 0.014 & 0.000 & 0.032 & 0.002 & -0.001 & -0.035 & 0.004\\
212.4--238.3 & $224.4$ &$(1.20 \, \pm 0.04 \, _{-0.06}^{+0.04}) \times 10^{-5}$ & 0.014 & 0.000 & 0.032 & 0.001 & 0.000 & -0.034 & 0.004\\
238.3--267.4 & $252.5$ &$(8.06 \, \pm 0.35 \, _{-0.39}^{+0.29}) \times 10^{-6}$ & 0.017 & 0.000 & 0.032 & -0.000 & 0.002 & -0.032 & 0.003\\
267.4--300.0 & $282.9$ &$(5.88 \, \pm 0.28 \, _{-0.27}^{+0.21}) \times 10^{-6}$ & 0.017 & 0.000 & 0.032 & -0.001 & 0.005 & -0.029 & 0.001\\
300.0--336.6 & $317.4$ &$(4.05 \, \pm 0.22 \, _{-0.18}^{+0.15}) \times 10^{-6}$ & 0.018 & 0.000 & 0.032 & -0.001 & 0.008 & -0.025 & -0.002\\
336.6--377.7 & $355.6$ &$(2.73 \, \pm 0.17 \, _{-0.11}^{+0.10}) \times 10^{-6}$ & 0.018 & 0.000 & 0.032 & -0.000 & 0.012 & -0.019 & -0.005\\
377.7--423.8 & $400.4$ &$(1.74 \, \pm 0.13 \, _{-0.07}^{+0.07}) \times 10^{-6}$ & 0.019 & 0.000 & 0.032 & 0.000 & 0.016 & -0.012 & -0.010\\
423.8--475.5 & $447.7$ &$(1.19 \, \pm 0.10 \, _{-0.05}^{+0.05}) \times 10^{-6}$ & 0.019 & 0.000 & 0.032 & 0.002 & 0.020 & -0.004 & -0.014\\
475.5--598.6 & $529.3$ &$(6.90 \, \pm 0.40 \, _{-0.44}^{+0.29}) \times 10^{-7}$ & $_{-0.051}^{+0.000}$ & 0.000 & 0.032 & 0.005 & 0.026 & 0.010 & -0.021\\
598.6--753.6 & $666.3$ &$(3.27 \, \pm 0.25 \, _{-0.22}^{+0.19}) \times 10^{-7}$ & $_{-0.053}^{+0.019}$ & 0.000 & 0.032 & 0.011 & 0.031 & 0.030 & -0.030\\
753.6--948.7 & $843.7$ &$(1.74 \, \pm 0.16 \, _{-0.17}^{+0.12}) \times 10^{-7}$ & $_{-0.083}^{+0.000}$ & 0.000 & 0.032 & 0.021 & 0.029 & 0.050 & -0.033\\
948.7--1194.3 & $1063.6$ &$(8.84 \, \pm 1.02 \, _{-0.57}^{+0.77}) \times 10^{-8}$ & $_{-0.051}^{+0.032}$ & 0.000 & 0.032 & 0.035 & 0.018 & 0.063 & -0.025\\
1194.3--1892.9 & $1463.2$ &$(2.04 \, \pm 0.30 \, _{-0.14}^{+0.23}) \times 10^{-8}$ & $_{-0.061}^{+0.066}$ & 0.000 & 0.032 & 0.059 & -0.015 & 0.062 & 0.010\\
1892.9--3000.0 & $2336.2$ &$(4.19 \, \pm 1.14 \, _{-0.85}^{+0.67}) \times 10^{-9}$ & $_{-0.183}^{+0.032}$ & 0.000 & 0.032 & 0.110 & -0.083 & 0.026 & 0.102\\
3000.0--4754.7 & $3815.3$ &$(9.36 \, _{-4.47}^{+5.37} \, _{-2.79}^{+2.11}) \times 10^{-10}$ & $_{-0.293}^{+0.064}$ & 0.000 & 0.032 & 0.187 & -0.045 & 0.090 & 0.051\\
\end{tabular}
\end{ruledtabular}
\end{table*}
\renewcommand{\arraystretch}{1.0}

\renewcommand{\arraystretch}{1.25}
\begin{table*}
\caption{Table of CALET electron plus positron spectrum using the same energy binning
as the DAMPE spectrum. The tabulated information is the same as the previous table. 
\label{tab:ele2}}
\begin{ruledtabular}
\begin{tabular}{cccrrrrrrr}
Energy Bin & Mean Energy & Flux & \multicolumn{7}{c}{Systematic Uncertainties (relative to flux)} \\
(GeV)  & (GeV) & (m$^{-2}$sr$^{-1}$s$^{-1}$GeV$^{-1}$) & $\sigma_{\rm BDT}$ & $\sigma_{\rm trig.}$ & $\Delta_{\rm norm.}$  & $\Delta_{\rm trk.}$ & $\Delta_{\rm chg.}$ & $\Delta_{\rm ID}$ & $\Delta_{\rm MC}$ \\
\hline
10.6--12.2 & $11.4$ &$(1.526 \, \pm 0.004 \, _{-0.103}^{+0.093}) \times 10^{-1}$ & 0.028 & 0.024 & 0.032 & -0.006 & -0.046 & 0.030 & 0.022\\
12.2--13.9 & $13.0$ &$(9.930 \, \pm 0.027 \, _{-0.439}^{+0.533}) \times 10^{-2}$ & 0.010 & 0.024 & 0.032 & 0.002 & -0.016 & 0.013 & 0.032\\
13.9--16.0 & $14.9$ &$(6.277 \, \pm 0.019 \, _{-0.258}^{+0.358}) \times 10^{-2}$ & 0.010 & 0.024 & 0.032 & 0.008 & 0.003 & -0.000 & 0.039\\
16.0--18.3 & $17.0$ &$(4.028 \, \pm 0.014 \, _{-0.171}^{+0.250}) \times 10^{-2}$ & 0.010 & 0.024 & 0.032 & 0.013 & 0.015 & -0.010 & 0.042\\
18.3--20.9 & $19.5$ &$(2.650 \, \pm 0.010 \, _{-0.118}^{+0.173}) \times 10^{-2}$ & 0.010 & 0.024 & 0.032 & 0.017 & 0.022 & -0.017 & 0.043\\
20.9--24.0 & $22.4$ &$(1.701 \, \pm 0.008 \, _{-0.079}^{+0.113}) \times 10^{-2}$ & 0.010 & 0.024 & 0.032 & 0.020 & 0.025 & -0.022 & 0.041\\
24.0--27.5 & $25.6$ &$(1.114 \, \pm 0.006 \, _{-0.053}^{+0.073}) \times 10^{-2}$ & 0.008 & 0.024 & 0.032 & 0.022 & 0.026 & -0.024 & 0.038\\
27.5--31.6 & $29.4$ &$(7.059 \, \pm 0.028 \, _{-0.339}^{+0.449}) \times 10^{-3}$ & 0.008 & 0.024 & 0.032 & 0.024 & 0.026 & -0.025 & 0.034\\
31.6--36.3 & $33.8$ &$(4.399 \, \pm 0.021 \, _{-0.186}^{+0.249}) \times 10^{-3}$ & 0.011 & 0.000 & 0.032 & 0.024 & 0.026 & -0.026 & 0.028\\
36.3--41.7 & $38.8$ &$(2.854 \, \pm 0.016 \, _{-0.119}^{+0.152}) \times 10^{-3}$ & 0.010 & 0.000 & 0.032 & 0.024 & 0.024 & -0.025 & 0.023\\
41.7--47.9 & $44.6$ &$(1.818 \, \pm 0.012 \, _{-0.075}^{+0.091}) \times 10^{-3}$ & 0.010 & 0.000 & 0.032 & 0.024 & 0.023 & -0.025 & 0.018\\
47.9--55.0 & $51.2$ &$(1.182 \, \pm 0.009 \, _{-0.049}^{+0.056}) \times 10^{-3}$ & 0.011 & 0.000 & 0.032 & 0.023 & 0.021 & -0.024 & 0.013\\
55.0--63.1 & $58.7$ &$(7.670 \, \pm 0.066 \, _{-0.318}^{+0.344}) \times 10^{-4}$ & 0.011 & 0.000 & 0.032 & 0.021 & 0.018 & -0.024 & 0.009\\
63.1--72.4 & $67.4$ &$(4.988 \, \pm 0.049 \, _{-0.206}^{+0.210}) \times 10^{-4}$ & 0.010 & 0.000 & 0.032 & 0.019 & 0.016 & -0.024 & 0.006\\
72.4--83.2 & $77.5$ &$(3.191 \, \pm 0.036 \, _{-0.135}^{+0.128}) \times 10^{-4}$ & 0.012 & 0.000 & 0.032 & 0.017 & 0.013 & -0.025 & 0.004\\
83.2--95.5 & $88.8$ &$(2.072 \, \pm 0.027 \, _{-0.089}^{+0.080}) \times 10^{-4}$ & 0.012 & 0.000 & 0.032 & 0.015 & 0.010 & -0.026 & 0.002\\
95.5--109.7 & $102.0$ &$(1.36 \, \pm 0.02 \, _{-0.06}^{+0.05}) \times 10^{-4}$ & 0.013 & 0.000 & 0.032 & 0.012 & 0.007 & -0.028 & 0.002\\
109.7--125.9 & $117.2$ &$(8.81 \, \pm 0.15 \, _{-0.40}^{+0.32}) \times 10^{-5}$ & 0.013 & 0.000 & 0.032 & 0.010 & 0.004 & -0.030 & 0.002\\
125.9--144.5 & $134.5$ &$(5.73 \, \pm 0.12 \, _{-0.27}^{+0.20}) \times 10^{-5}$ & 0.014 & 0.000 & 0.032 & 0.008 & 0.002 & -0.032 & 0.003\\
144.5--166.0 & $154.3$ &$(3.80 \, \pm 0.09 \, _{-0.19}^{+0.14}) \times 10^{-5}$ & 0.014 & 0.000 & 0.032 & 0.005 & -0.000 & -0.034 & 0.004\\
166.0--190.6 & $177.6$ &$(2.54 \, \pm 0.07 \, _{-0.13}^{+0.09}) \times 10^{-5}$ & 0.014 & 0.000 & 0.032 & 0.003 & -0.001 & -0.035 & 0.004\\
190.6--218.8 & $203.7$ &$(1.63 \, \pm 0.05 \, _{-0.08}^{+0.06}) \times 10^{-5}$ & 0.014 & 0.000 & 0.032 & 0.002 & -0.001 & -0.035 & 0.004\\
218.8--251.2 & $233.7$ &$(1.03 \, \pm 0.04 \, _{-0.05}^{+0.04}) \times 10^{-5}$ & 0.014 & 0.000 & 0.032 & 0.000 & 0.001 & -0.034 & 0.003\\
251.2--288.4 & $268.5$ &$(6.88 \, \pm 0.28 \, _{-0.33}^{+0.25}) \times 10^{-6}$ & 0.017 & 0.000 & 0.032 & -0.001 & 0.003 & -0.031 & 0.002\\
288.4--331.1 & $308.2$ &$(4.51 \, \pm 0.21 \, _{-0.20}^{+0.17}) \times 10^{-6}$ & 0.018 & 0.000 & 0.032 & -0.001 & 0.007 & -0.026 & -0.001\\
331.1--380.2 & $353.3$ &$(2.85 \, \pm 0.16 \, _{-0.12}^{+0.11}) \times 10^{-6}$ & 0.018 & 0.000 & 0.032 & -0.000 & 0.011 & -0.020 & -0.005\\
380.2--436.5 & $407.5$ &$(1.72 \, \pm 0.11 \, _{-0.07}^{+0.07}) \times 10^{-6}$ & 0.019 & 0.000 & 0.032 & 0.001 & 0.017 & -0.011 & -0.010\\
436.5--501.2 & $466.5$ &$(1.12 \, \pm 0.07 \, _{-0.05}^{+0.05}) \times 10^{-6}$ & 0.019 & 0.000 & 0.032 & 0.002 & 0.022 & -0.001 & -0.016\\
501.2--575.4 & $536.5$ &$(6.50 \, \pm 0.50 \, _{-0.42}^{+0.28}) \times 10^{-7}$ & $_{-0.051}^{+0.000}$ & 0.000 & 0.032 & 0.005 & 0.026 & 0.011 & -0.022\\
575.4--660.7 & $615.3$ &$(4.49 \, \pm 0.38 \, _{-0.30}^{+0.24}) \times 10^{-7}$ & $_{-0.053}^{+0.019}$ & 0.000 & 0.032 & 0.009 & 0.030 & 0.023 & -0.027\\
660.7--758.6 & $708.9$ &$(2.58 \, \pm 0.28 \, _{-0.18}^{+0.16}) \times 10^{-7}$ & $_{-0.053}^{+0.019}$ & 0.000 & 0.032 & 0.014 & 0.031 & 0.036 & -0.032\\
758.6--871.0 & $809.7$ &$(1.87 \, \pm 0.22 \, _{-0.18}^{+0.13}) \times 10^{-7}$ & $_{-0.083}^{+0.000}$ & 0.000 & 0.032 & 0.019 & 0.030 & 0.047 & -0.033\\
871.0--1000.0 & $929.9$ &$(1.38 \, \pm 0.18 \, _{-0.13}^{+0.10}) \times 10^{-7}$ & $_{-0.083}^{+0.000}$ & 0.000 & 0.032 & 0.026 & 0.026 & 0.056 & -0.031\\
1000.0--1148.2 & $1080.5$ &$(9.31 \, \pm 1.35 \, _{-0.60}^{+0.81}) \times 10^{-8}$ & $_{-0.051}^{+0.032}$ & 0.000 & 0.032 & 0.036 & 0.017 & 0.063 & -0.024\\
1148.2--1318.3 & $1224.8$ &$(4.48 \, \pm 0.89 \, _{-0.27}^{+0.41}) \times 10^{-8}$ & $_{-0.051}^{+0.032}$ & 0.000 & 0.032 & 0.044 & 0.006 & 0.065 & -0.013\\
1318.3--1513.6 & $1400.8$ &$(1.67 \, \pm 0.52 \, _{-0.12}^{+0.19}) \times 10^{-8}$ & $_{-0.061}^{+0.066}$ & 0.000 & 0.032 & 0.055 & -0.009 & 0.064 & 0.004\\
1513.6--1737.8 & $1618.3$ &$(1.88 \, \pm 0.52 \, _{-0.14}^{+0.22}) \times 10^{-8}$ & $_{-0.061}^{+0.066}$ & 0.000 & 0.032 & 0.069 & -0.030 & 0.057 & 0.028\\
1737.8--1995.3 & $1891.3$ &$(1.25 \, \pm 0.39 \, _{-0.11}^{+0.17}) \times 10^{-8}$ & $_{-0.061}^{+0.066}$ & 0.000 & 0.032 & 0.085 & -0.053 & 0.045 & 0.059\\
1995.3--2290.9 & $2162.0$ &$(3.79 \, _{-2.04}^{+3.08} \, _{-0.76}^{+0.55}) \times 10^{-9}$ & $_{-0.183}^{+0.032}$ & 0.000 & 0.032 & 0.100 & -0.073 & 0.032 & 0.087\\
2290.9--2630.3 & $2481.2$ &$(3.06 \, _{-1.37}^{+2.29} \, _{-0.63}^{+0.52}) \times 10^{-9}$ & $_{-0.183}^{+0.032}$ & 0.000 & 0.032 & 0.118 & -0.089 & 0.022 & 0.111\\
2630.3--3019.9 & $2873.3$ &$(3.13 \, _{-1.63}^{+2.04} \, _{-0.65}^{+0.60}) \times 10^{-9}$ & $_{-0.183}^{+0.032}$ & 0.000 & 0.032 & 0.140 & -0.096 & 0.020 & 0.124\\
3019.9--3467.4 & $3133.1$ &$(1.05 \, _{-0.71}^{+1.43} \, _{-0.32}^{+0.22}) \times 10^{-9}$ & $_{-0.293}^{+0.064}$ & 0.000 & 0.032 & 0.153 & -0.091 & 0.027 & 0.119\\
3467.4--3981.1 & $3738.3$ &$(9.60 \, _{-6.45}^{+12.67} \, _{-2.88}^{+2.12}) \times 10^{-10}$ & $_{-0.293}^{+0.064}$ & 0.000 & 0.032 & 0.183 & -0.053 & 0.079 & 0.063\\
3981.1--4570.9 & $4170.3$ &$(8.29 \, _{-5.60}^{+11.21} \, _{-2.45}^{+2.17}) \times 10^{-10}$ & $_{-0.293}^{+0.064}$ & 0.000 & 0.032 & 0.203 & -0.002 & 0.149 & -0.020\\
\end{tabular}
\end{ruledtabular}
\end{table*}
\renewcommand{\arraystretch}{1.0}
\clearpage

\begin{figure}[bth!]
\begin{center}
\includegraphics[width=\hsize]{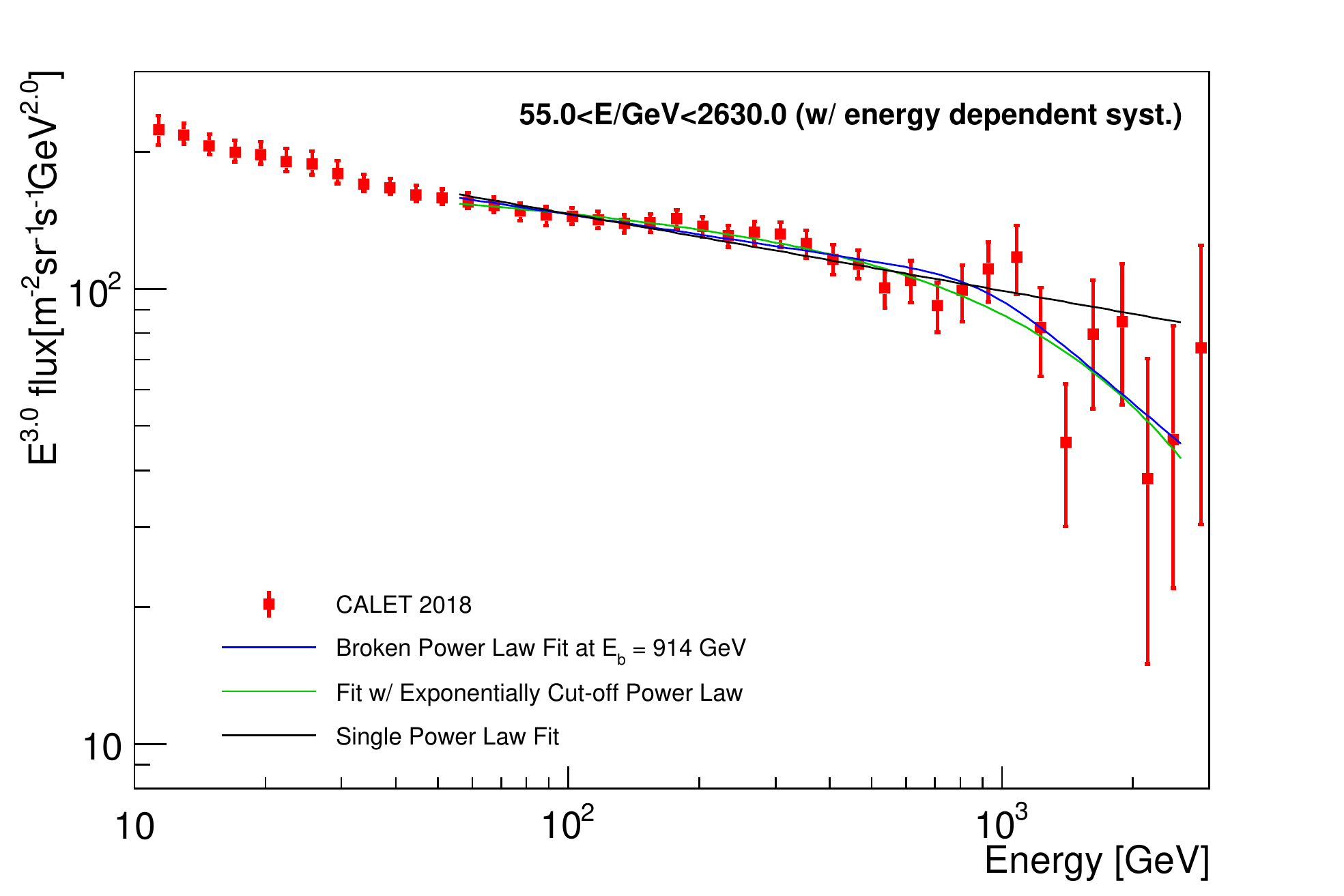}
\caption{The fit of the CALET all-electron spectrum with a smoothly broken power law model~\cite{SM_DAMPE2017}
(blue line),
while fixing the break energy at 914~GeV as determined by DAMPE~\cite{SM_DAMPE2017}.
The smoothly broken power law model is defined as: 
$\Phi(E) = \Phi_{0} (E/100~{\rm GeV})^{-\gamma_{1}} [1 + (E/E_{b})^{-(\gamma_{1} - \gamma_{2})/\Delta}]^{-\Delta}$, 
where $E_{b}$ is the break energy, while $\gamma_{1}$ and $\gamma_{2}$ are 
the power law indices below and above the
break energy, respectively. The smoothness parameter, $\Delta$, is fixed to 0.1.
The fitting yields $\gamma_{1} = -3.15 \pm 0.02$ and $\gamma_{2} = -3.81 \pm 0.32$
with a $\chi^{2}$ of 17.0 and number of degree of freedom (NDF) being 25.
If we fit the spectrum with an exponentially cut-off power law~\cite{SM_Fermi2017-e}
(green line),
we obtain 2.3$\pm$0.7~TeV as the cutoff energy and a spectral index
of $-3.06 \pm 0.03$ with 
$\chi^{2}/NDF = 13.0/25$.
On the other hand, a single power law fit (black line) in the same energy range gives 
an index of $-3.17 \pm 0.02$ with $\chi^{2}/{\rm NDF} = 26.5/26$.
All the parameters are consistent within errors between this energy binning (as shown in Table~\ref{tab:ele2}) 
and our original energy binning (as shown in Table~\ref{tab:ele}). 
}
\label{fig:bpl}
\end{center}
\end{figure}
\clearpage

\begin{figure}[bth!]
\begin{center}
\includegraphics[width=\hsize]{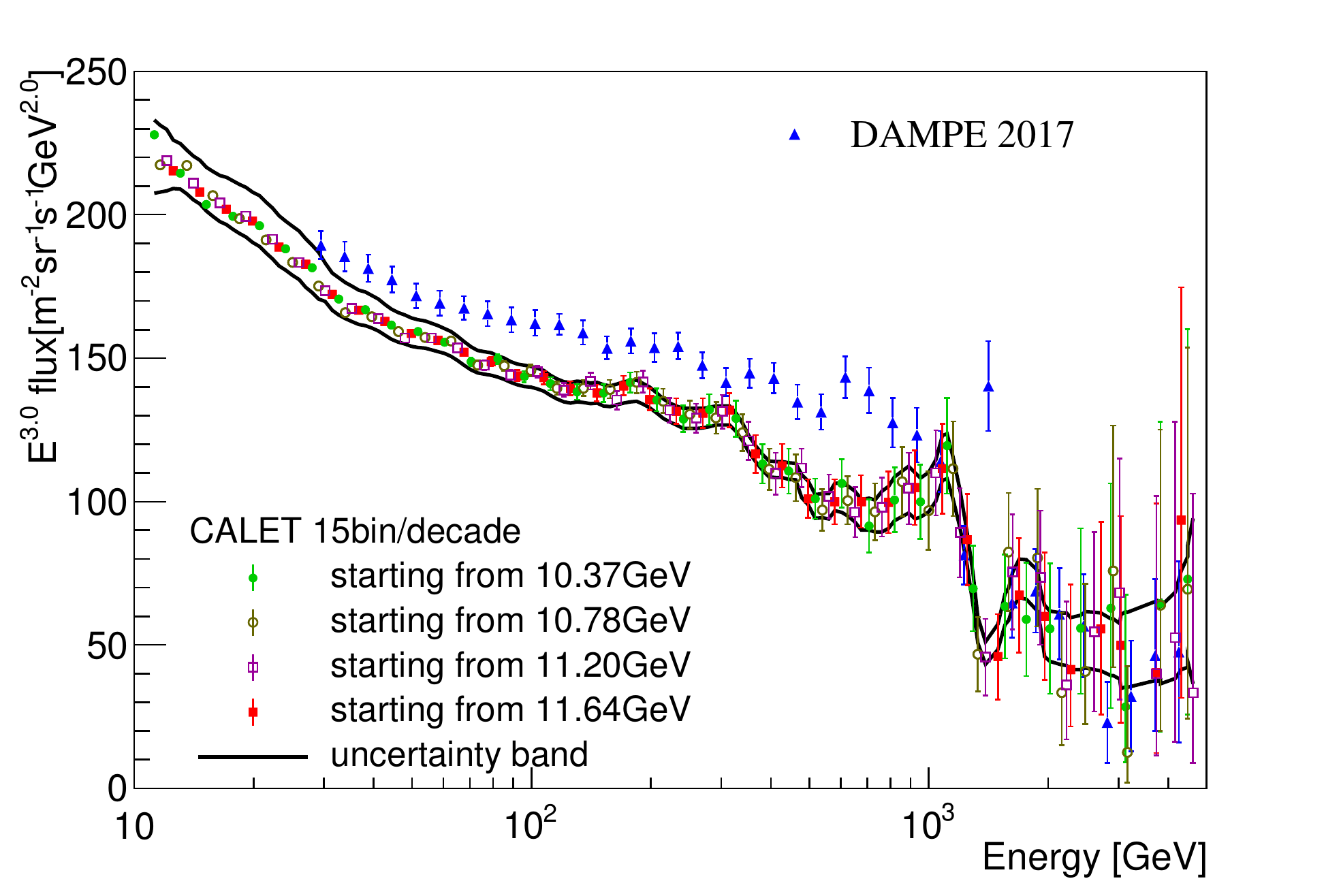}
\caption{
Study of possible binning related effects in the CALET electron and positron spectrum. 
To further test the presence of potential systematics in our spectrum, 
different choices for the adopted binning are checked 
where the bin-width is chosen as 15~bins/decade in equal log-bins and four different
binnings are shown in the same plot. 
Each binning is shifted by one fourth of the 
bin width to study possible binning related issues.
The solid curves in the figure show the energy dependent systematic uncertainty band.
Transition from event selection A+B with K-cut to A+B+C+D acceptance with BDT
occurs in both cases at the low-edge of the bin containing 500~GeV. 
As demonstrated, the deviation due to binning is well
below our energy dependent systematic uncertainty or statistical fluctuations.
Therefore, bin-to-bin migration and related effects are negligible compared to 
our estimated systematic uncertainties,
in accordance with the estimated CALET energy resolution of 2\% above 20~GeV. 
}
\label{fig:binning15}
\end{center}
\end{figure}

\providecommand{\noopsort}[1]{}\providecommand{\singleletter}[1]{#1}%

\end{document}